\documentclass[useAMS,usenatbib]{mn2e}
\input psfig.sty

\def \aj {AJ}
\def \apj {ApJ}
\def \apjl {ApJL}
\def \mnras {MNRAS}
\def \apjs {ApJS}
\def \aap {A\&A}
\def \nat {Nature}
\def \jcap {JCAP}
\def \etal {et~al.~}

\def \spose#1{\hbox  to 0pt{#1\hss}}  
\def \lta{\mathrel{\spose{\lower 3pt\hbox{$\sim$}}\raise  2.0pt\hbox{$<$}}}
\def \gta{\mathrel{\spose{\lower  3pt\hbox{$\sim$}}\raise 2.0pt\hbox{$>$}}}

\def \kmsmpc {\>{\rm km}\,{\rm s}^{-1}\,{\rm Mpc}^{-1}}
\def \kms {\ifmmode  \,\rm km\,s^{-1} \else $\,\rm km\,s^{-1}  $ \fi }
\def \kpc {\ifmmode  {\rm kpc}  \else ${\rm  kpc}$ \fi  }  
\def \Msun {\ifmmode {\rm M}_{\odot} \else ${\rm M}_{\odot}$ \fi} 
\def \Mstar {\ifmmode M_{\rm star} \else $M_{\rm star}$ \fi} 
\def \Mtot {\ifmmode M_{\rm tot} \else $M_{\rm tot}$ \fi} 
\def \Msps {\ifmmode M_{\rm SPS} \else $M_{\rm SPS}$ \fi} 
\def \Mdyn {\ifmmode M_{\rm dyn} \else $M_{\rm dyn}$ \fi} 
\def \Re {\ifmmode R_{\rm e} \else $R_{\rm e}$ \fi} 
\def \sigmaap {\ifmmode \sigma_{\rm ap} \else $\sigma_{\rm ap}$ \fi} 
\def \sigmae  {\ifmmode \sigma_{\rm e}  \else $\sigma_{\rm e}$ \fi} 
\def \sigmaei {\ifmmode \sigma_{\rm e8} \else $\sigma_{\rm e8}$ \fi}

\def \LCDM {\ifmmode \Lambda{\rm CDM} \else $\Lambda{\rm CDM}$ \fi}
\def \OmegaM {\ifmmode \Omega_{\rm M} \else $\Omega_{\rm M}$ \fi} 
\def \OmegaL {\ifmmode \Omega_{\rm \Lambda} \else $\Omega_{\rm \Lambda}$\fi} 

\def \Mvir {\ifmmode M_{\rm  vir} \else $M_{\rm  vir}$ \fi}  

\def \DeltaIMF {\ifmmode \Delta_{\rm IMF} \else $\Delta_{\rm IMF}$ \fi}
\def \dvr {\ifmmode \partial_{\rm VR} \else $\partial_{\rm VR}$ \fi}

\title[IMF and the Fundamental Plane] {Universal IMF {\it vs} dark
  halo response in early-type galaxies: breaking the degeneracy with
  the Fundamental Plane}

\author[Dutton et al.]  {Aaron  A. Dutton$^{1,2}$\thanks{dutton@mpia.de}, 
  Andrea V. Macci\`o$^1$, J. Trevor Mendel$^2$, Luc Simard$^3$\\
  $^1$Max Planck Institute for Astronomy, K\"onigstuhl 17, 69117,
      Heidelberg, Germany\\
  $^2$Dept. of Physics and Astronomy, University of Victoria, Victoria, 
      B.C., V8P 5C2, Canada\\
  $^3$Herzberg Institute of Astrophysics, National Research Council of
      Canada, 5071 West Saanich road, Victoria, B.C., V9E 2E7, Canada\\}
\begin{document}
             
\date{Accepted 2013 April 10. Received 2013 April 3; in original form
  2012 April 12}
             
\pagerange{\pageref{firstpage}--\pageref{lastpage}}\pubyear{2013}

\maketitle           

\label{firstpage}
             

\begin{abstract}
  We use the relations between aperture stellar velocity dispersion
  ($\sigmaap$), stellar mass ($\Msps$), and galaxy size $(\Re$) for a
  sample of $\sim 150\,000$ early-type galaxies from SDSS/DR7 to place
  constraints on the stellar initial mass function (IMF) and dark halo
  response to galaxy formation.  We build \LCDM based mass models that
  reproduce, by construction, the relations between galaxy size, light
  concentration and stellar mass, and use the spherical Jeans
  equations to predict $\sigmaap$. Given our model assumptions
  (including those in the stellar population synthesis models), we
  find that reproducing the median $\sigmaap$ vs $\Msps$ relation is
  not possible with {\it both} a universal IMF and a universal dark
  halo response. Significant departures from a universal IMF and/or
  dark halo response are required, but there is a degeneracy between
  these two solutions.
  We show that this degeneracy can be broken using the strength of the
  correlation between residuals of the velocity-mass
  ($\Delta\log\sigmaap$) and size-mass ($\Delta\log\Re$)
  relations. The slope of this correlation, $\dvr\equiv
  \Delta\log\sigmaap/\Delta\log\Re$, varies systematically with galaxy
  mass from $\dvr\simeq -0.45$ at $\Msps\sim 10^{10}\Msun$, to
  $\dvr\simeq -0.15$ at $\Msps\sim 10^{11.6}\Msun$.  The virial
  fundamental plane (FP) has $\dvr=-1/2$, and thus we find the tilt of
  the observed FP is mass dependent.
  Reproducing this tilt requires {\it both} a non-universal IMF and a
  non-universal halo response. Our best model has mass-follows-light
  at low masses ($\Msps \lta 10^{11.2}\Msun$) and unmodified NFW
  haloes at $\Msps\sim 10^{11.5}\Msun$.  The stellar masses imply a
  mass dependent IMF which is ``lighter'' than Salpeter at low masses
  and ``heavier'' than Salpeter at high masses.
\end{abstract}


\setcounter{footnote}{1}


\section{Introduction}
\label{sec:intro}

The form of the stellar initial mass function (IMF) and the response
of dark matter haloes to galaxy formation are two fundamental
unknowns, which are important in many areas of astrophysics. For
example, the IMF is needed in order to convert observations of
integrated stellar light into stellar masses and star formation rates
(two fundamental parameters in galaxy evolution studies), and for
modelling the supernova rates, chemical evolution, and production of
ionizing photons in galaxies.
The response of dark matter haloes to galaxy formation is needed in
order to constrain the nature of dark matter from observations of the
structure of dark matter haloes (e.g., central density slopes, central
densities) and indirect dark matter detection experiments. Halo
response also plays a key role in determining the fundamental scaling
relations between velocity and light of galaxies (Faber \& Jackson
1976, Tully \& Fisher 1977).

For many years the IMF and dark halo response were thought to be
universal: The IMFs of external galaxies are the same as measured in
the Milky Way (Kroupa 2001, Chabrier 2003); and dark matter haloes
contract adiabatically in response to galaxy formation (Blumenthal
\etal 1986; Gnedin \etal 2004). However, recent observations and
numerical simulations have cast doubt on these assumptions. We note
that while there have been claims of observational evidence for dark
halo contraction in early-type galaxies (e.g., Schulz \etal 2010;
Napolitano \etal 2010; Trujillo-Gomez \etal 2011; Chae \etal 2012)
these conclusions depend (trivially) on the assumption of a universal
Milky Way type IMF (e.g., Dutton \etal 2011a).

It has been known for many years that baryonic effects can, in
principle, result in reduced halo contraction or even halo expansion
(e.g., Navarro, Eke, \& Frenk 1996; El-Zant \etal 2001). But only
recently have these effects been demonstrated in fully cosmological
simulations of galaxy formation (Johansson \etal 2009; Abadi \etal
2010; Duffy \etal 2010; Governato \etal 2010; Macci\`o \etal 2012;
Martizzi \etal 2012). Although some authors still maintain that dark
halo contraction is an unavoidable consequence of galaxy formation in
\LCDM cosmologies (e.g., Gnedin \etal 2011).

The traditional route to constraining the IMF is through direct star
counts. In the Milky Way this has shown the IMF to have a power-law
shape $dN/dm \propto m^{-x}$, with $x\simeq -2.3$ at masses above
$m\simeq 1\Msun$ (Salpeter 1955), and a turn over at lower masses
(Kroupa 2001; Chabrier 2003). Since most of the mass is in low-mass
stars, Kroupa/Chabrier IMFs have lower stellar mass-to-light ratios
(by about 0.20 to 0.25 dex) than a Salpeter IMF.  Outside of our Galaxy,
counting individual stars is usually not feasible (especially down to
the low masses required to fully constrain the IMF). There are several
approaches which can be used to probe the IMF at different mass scales
in extragalactic systems.

Constraints on the shape of the IMF at high masses typically involve
measurements of H$\alpha$ and/or far ultraviolet (FUV) fluxes. For
example, the equivalent width of H$\alpha$ can constrain the slope of
the IMF above $\sim 1\Msun$ (Kennicutt 1983, Hoversten \& Glazebrook
2008), and the ratio between H$\alpha$ and FUV fluxes probes the IMF above
$\sim 10\Msun$ (Meurer \etal 2009; Lee \etal 2009). Broadly speaking
these methods show that the high mass IMF in external galaxies is
similar to that of the Milky Way, but there are systematic
discrepancies which could be explained by variation in the slope
and/or upper mass limit of the IMF. Note that while modest changes in
the upper end of the IMF don't impact the stellar mass-to-light ratios
of old stellar populations, top heavy IMFs can leave behind enough
stellar remnants to significantly increase stellar mass-to-light
ratios.

The low mass end (below $\sim 1\Msun$) of the IMF can be probed by
comparing the mass (or an upper limit) of all the stars from dynamics
and/or strong gravitational lensing to the stellar mass obtained from
stellar population synthesis (SPS) models. Such studies require IMFs
lighter (i.e., lower stellar mass-to-light ratios) than Salpeter in
spiral galaxies (Bell \& de Jong 2001; Bershady \etal 2011; Dutton
\etal 2011a, Barnab{\`e} \etal 2012; Brewer \etal 2012), but are
consistent with a universal Milky Way type IMF.

In elliptical and lenticular galaxies, Cappellari \etal (2006) showed
that a Salpeter IMF over-predicts the dynamical mass-to-light ratios
of some ``fast-rotating'' galaxies.  Using dynamics and strong
gravitational lensing from the SLACS survey (Bolton \etal 2006) Treu
\etal (2010) showed that in fact a Salpeter IMF is allowed in massive
(stellar mass $\Mstar \gta 10^{11}\Msun$, velocity dispersion $\sigma
\gta 200 \kms$) elliptical galaxies, even accounting for ``standard''
dark matter haloes. Extending this study to include constraints from
weak lensing, predictions from \LCDM, and the possibility of dark halo
contraction, Auger \etal (2010a) concluded that a Salpeter-type IMF
was strongly favored over Milky Way type IMF.  In order to minimize
uncertainties in subtracting off the dark matter, Dutton, Mendel, \&
Simard (2012) studied the densest $\simeq 3\%$ of early-type galaxies
in the SDSS. These galaxies have a fundamental plane correlation
consistent with no dark matter within an effective radius. The average
IMF of these galaxies is close to Salpeter, with evidence that
redder/bluer galaxies have heavier/lighter IMFs. A Salpeter-type IMF
is also favored in the bulges of massive spiral galaxies (Dutton \etal
2013), and brightest cluster galaxies (Newman \etal 2013).

Another method for probing the low mass end of the IMF (in non star
forming galaxies) is by measuring the strength of dwarf star sensitive
absorption lines.  Applying this method to nearby massive elliptical
galaxies van Dokkum \& Conroy (2010) found an IMF steeper (below $\sim
1 \Msun$) than that of the Milky Way, and most likely steeper than
Salpeter. This result has been confirmed with larger samples of
galaxies (Smith \etal 2012; Spiniello \etal 2012; Conroy \& van Dokkum
2012). These studies also find evidence for a dependence of the (low
mass) IMF slope on stellar velocity dispersion and $\alpha$-abundance,
with steeper slopes at higher dispersions and shorter star formation
timescales.

Thus there is good agreement from different methods that the IMFs of
the most massive or most dense elliptical galaxies are heavier than
that of the Milky Way, and most likely similar to a Salpeter IMF below
$\sim 1\Msun$.  The next step is to constrain how the IMF varies in a
more general sample of early-type galaxies.  The ATLAS3D project has
made progress towards this aim (Cappellari \etal 2012a), but their
study is limited by the inherent degeneracies between the baryonic and
dark matter components in the mass modeling of galaxies (e.g., Dutton
\etal 2005). Specifically there is a factor of $\sim 2$ range in
derived stellar mass-to-light ratios between models in which
mass-follows-light and models with cosmologically motivated
adiabatically contracted dark matter haloes (see also Dutton \etal
2011a). 

In this paper we place constraints on the mass dependence of the IMF
and dark halo response using scaling relations of a large sample
($>10^5$) of early-type galaxies from the Sloan Digital Sky Survey
(SDSS, York \etal 2000). This approach has strengths and weaknesses
compared to detailed mass models of individual galaxies (e.g., Dutton
\etal 2011b; Sonnenfeld \etal 2012; Cappellari \etal 2012a). The
weakness is we only get constraints for average galaxies (of a given
mass), the strength is that our constraints are free of statistical
uncertainties (in e.g., stellar mass-to-light ratios, dark halo
masses, dark halo concentrations, etc) that necessarily effect studies
with smaller samples. Both approaches are valid, providing
complementary constraints to the variability of the IMF and dark halo
response.

We construct dynamical models that consist of spherical distributions
of stars and dark matter.  These models are constrained to reproduce a
number of observational and theoretical constraints (e.g., Dutton
\etal 2011a). These models have two unknowns: the stellar IMF and the
response of the dark matter halo to galaxy formation. We use the
observed stellar velocity dispersion - stellar mass (Faber-Jackson)
relation to single out allowable combinations of IMF and halo
response. We show that the IMF and/or the halo response has to vary
with galaxy mass. However, from the velocity dispersion vs stellar
mass relation alone, we are unable to uniquely determine which if
either should vary, or the absolute normalization of the IMF and halo
response.

Our additional constraint is the slope of the correlation between the
residuals of the velocity - stellar mass ($\Delta\log\sigma$) and size
- stellar mass ($\Delta\log R$) relations, which for brevity we denote
$\dvr \equiv \Delta\log \sigma / \Delta \log R$.  This is a useful
constraint because it depends on the dark matter fraction within
(roughly) the half-light radius (Courteau \& Rix 1999; Dutton \etal
2007).  If there is no dark matter then $\dvr=-1/2$, while dark matter
dominated galaxies are expected to have $\dvr > 0$.  As we show below
$\dvr$ is related to a more familiar concept -- namely the tilt of the
fundamental plane (FP, Dressler \etal 1987; Djorgovski \& Davis 1987).
A related approach was taken by Borriello \etal (2003), who used the
fundamental plane constraints the dark matter content of early-type
galaxies.

This paper is organized as follows. The relation between $\dvr$ and
the tilt of the FP is shown in \S2, the main observed scaling
relations are presented in \S3, with additional details given in
appendix A. Constraints on the IMF and halo response from the velocity
- mass relation and the tilt of the FP are discussed in \S4, and \S5
respectively. A discussion of systematic effects is given in \S6, and
a summary is given in \S7. Appendix A gives the details of our mass
models. We adopt a \LCDM cosmology with $\OmegaL=0.7$, $\OmegaM=0.3$
and $H_0=70 \kmsmpc$

\section{The tilt of the Fundamental Plane}
\label{sec:fp}

The fundamental plane is a well-known scaling relation between the
size ($R$), and surface brightness ($I$), and velocity dispersion
($\sigma$) of early-type galaxies:
\begin{equation}
\label{eq:fp1}
R\propto \sigma^a I^b.
\end{equation}
For $r$-band sizes and surface brightnesses, the exponents are
$(a,b)=(1.49\pm0.05,-0.75\pm 0.01)$ (e.g. Bernardi \etal 2003).  This
differs from the simplest mass-follows-light model, known as the
virial plane, in which $(a,b)=(2,-1)$. The differences in exponent are
referred to as the tilt of the fundamental plane.

The origin of the tilt has been the subject of debate for many years
(e.g., van der Marel 1991; Bender, Burstein \& Faber 1992; Ciotti,
Lanzoni, \& Renzini 1996; Graham \& Colless 1997; Prugniel \& Simien
1996, 1997; Pahre, Djorgovski, \& de Carvalho 1998; Padmanabhan \etal
2004; Trujillo, Burkert, \& Bell 2004; Robertson \etal 2006; Bolton
\etal 2007, 2008; Graves \& Faber 2010). There are two principle
explanations for the tilt: (1) Variation of total mass to light ratio;
and (2) Structural non-homology.
Using the fact that $I\propto L/R^2$ one can re-write Eq.~\ref{eq:fp1} as
\begin{equation}
R\sigma^2 / L \propto \sigma^{2-a} I^{-1-b}.
\end{equation}
Thus the virial plane corresponds to $R\sigma^2/L=$ constant, and the
tilt of the fundamental plane corresponds to variation in
$R\sigma^2/L$.

In the case of structural homology the total mass, $\Mtot$, within
radius $R$, is directly proportional to $R\sigma^2$, and thus
variation in $R\sigma^2/L$ corresponds to variation in $\Mtot/L$.  The
total mass-to-light ratio can vary as a result of variation in stellar
mass-to-light ($\Mstar/L$) or variation in the total mass to stellar
mass $(\Mtot/\Mstar)$ ratio. The former is a result of stellar
population differences, while the latter is a result of variation of
dark matter fractions.

In the case of constant $\Mtot/L$, structural non-homology results in
the total mass not being directly proportional to
$R\sigma^2$. Examples of structural non-homology are variation of the
stellar mass profile (e.g., a variation in the S\'ersic index or
concentration parameter), and variation in the anisotropy of stellar
orbits.

Going back to the fundamental plane as written Eq.~\ref{eq:fp1}, both
coefficients are tilted with respect to the virial prediction. It
turns out that the fundamental plane can be written so that the tilt
is in a single coefficient. We can re-write Eq.~\ref{eq:fp1} as
\begin{equation}
\label{eq:fp2}
\sigma \propto L^{c} R^{d}
\end{equation}
where $c=-b/a$, and $d=(1+2b)/a$. The corresponding virial
coefficients are $(c,d)=(1/2,-1/2)$.  The exponents from Bernardi
\etal (2003) result in $(c,d)=(0.50\pm 0.02,-0.33\pm 0.01)$, and thus
the tilt of the $r$-band fundamental plane is consistent with being
entirely due to $d\ne -1/2$.
Replacing luminosity with stellar mass in Eq.~\ref{eq:fp2}, we see
that the strength of the correlation between the velocity-mass and
size-mass relations is equivalent to the tilt of the fundamental
plane, i.e., $\dvr \equiv d$.

\begin{figure*}
\centerline{
\psfig{figure=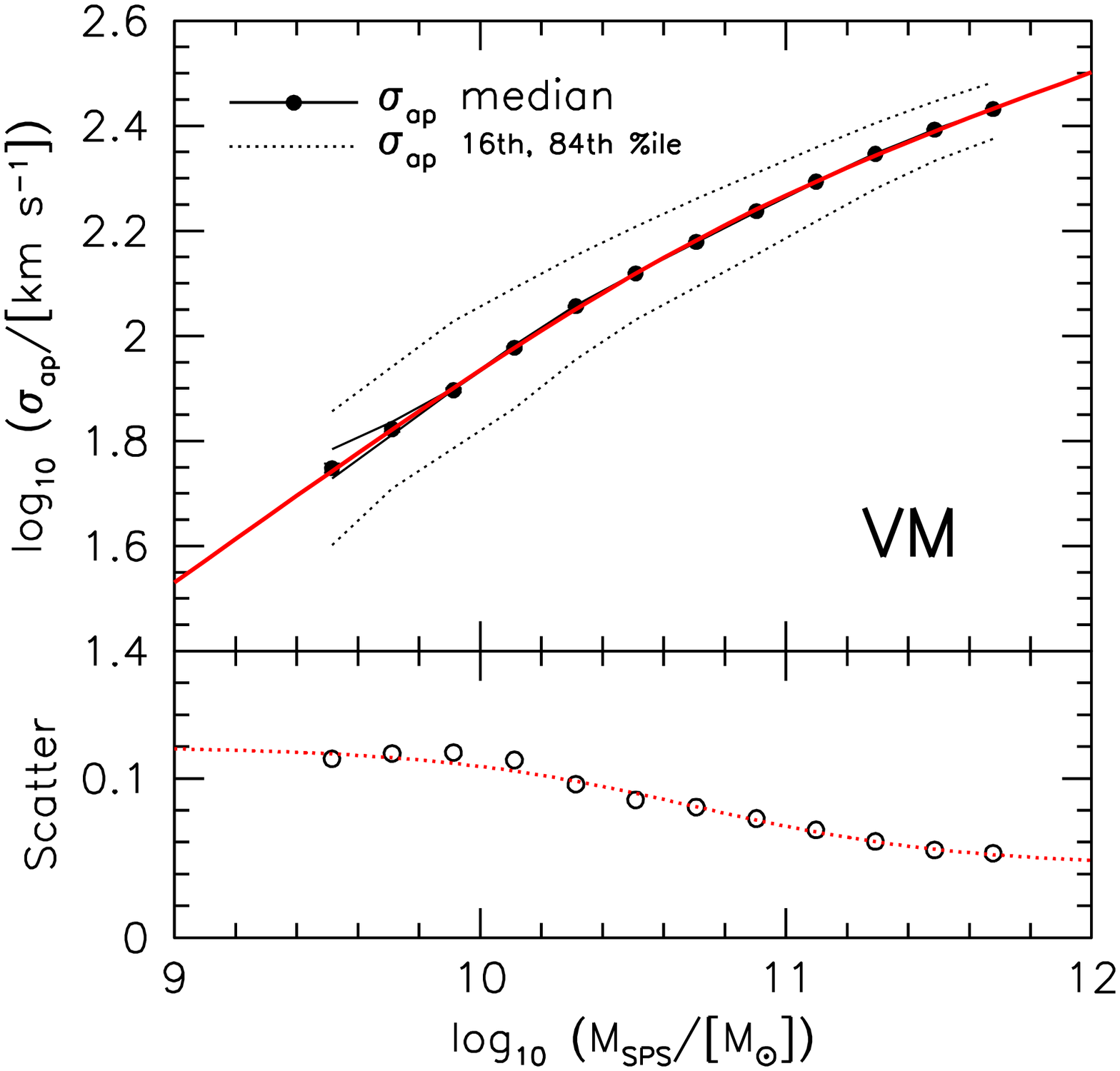,width=0.49\textwidth}
\psfig{figure=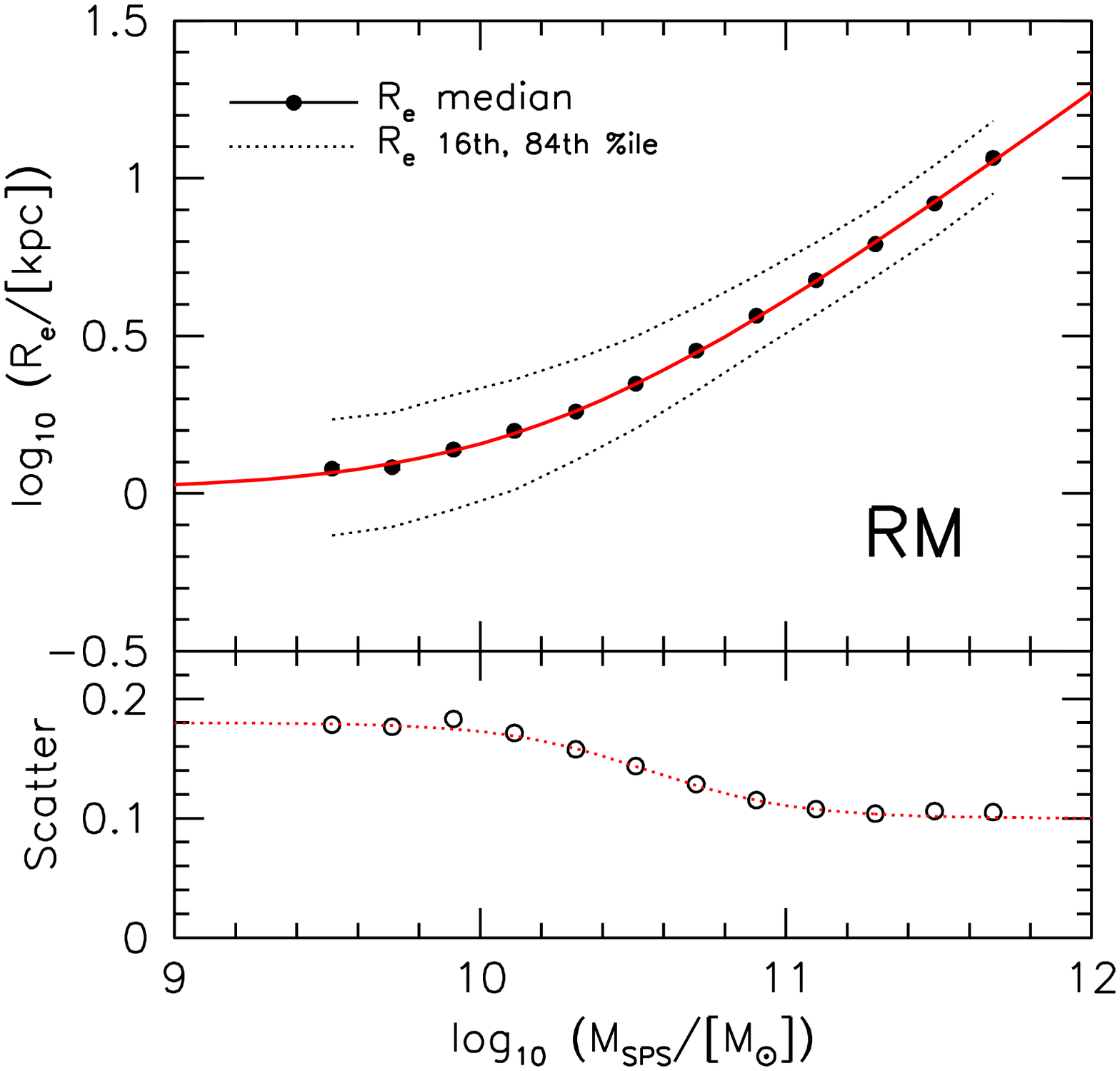,width=0.49\textwidth}
}
\caption{Observed velocity - stellar mass (VM, left) and size -
  stellar mass (RM, right) relations for SDSS early-type galaxies.
  Velocity dispersions ($\sigma_{\rm ap}$) are measured within the
  SDSS 3 arcsec diameter aperture, the stellar masses ($M_{\rm SPS}$)
  assume a Chabrier (2003) IMF, the sizes ($\Re$) are circularized
  half-light radii measured in the $r-$band. For the VM relation the
  black solid lines show the relations obtained for the two different
  velocity dispersion measurements, which are only visibly different
  at low masses. {\it Upper panels:} The solid points show medians in
  bins of 0.2 dex width in $\Msps$. The solid red lines show double
  power-law fits to these points (parameters in Table~\ref{tab:fits}),
  while the dotted lines show the 16th and 84th percentiles. {\it
    Lower panel:} The open points show the scatter in each mass
  bin. The red dotted lines show fits to these points (parameters
  given in Table~\ref{tab:fits}). }
\label{fig:vmr}
\end{figure*}

\section{Observed Scaling Relations}
\label{sec:obs}

\subsection{Sample Overview}
Here we give an overview of our observational sample of early-type
galaxies. We use a similar (but not identical) selection procedure as
described in more detail in Dutton \etal (2011a).  In summary we use
four selection criteria: (1) A spectroscopic redshift $0.005 < z <
0.3$; (2) A spectrum classified as early-type by SDSS (eCLASS $< 0$);
(3) A red $(g-r)$ color, based on valley in color - stellar mass
plane; (4) A minor-to-major axis ratio, $b/a > 0.5$ (which removes
dusty edge-on spirals).  These cuts are designed to select non-star
forming galaxies, so our sample includes ellipticals and lenticulars.
We also apply a redshift dependent minimum stellar mass to remove the
color bias at fixed stellar mass which results from the SDSS $r$-band
magnitude limit for spectroscopy. Our final sample consists of $\sim
150 \,000$ early-type galaxies.

We use structural parameters (such as sizes and axis ratios) from the
S\'ersic $n = 4$ plus $n = 1$ fits of Simard \etal (2011). Unless
otherwise stated the sizes we use here, $R_{\rm e}$, are the
circularized half-light sizes derived from the model of the total
light profile.

There are two velocity dispersion measurements publicly available for
SDSS galaxies. We refer to these as ``SDSS'' and ``Princeton''. At
most masses these two measurements are in excellent agreement, in
addition the scatter in these two measurements is in good agreement
with the reported uncertainties. However, at low masses (and low
dispersions) there are significant differences (see also Hyde \&
Bernardi 2009). In what follows, for the observed velocity dispersion
we adopt the logarithmic average of the two measurements. Where
appropriate we will use the individual measurements to gauge
systematic uncertainties in scaling relations.

When constructing the Faber-Jackson relation and the fundamental plane
from fiber spectroscopy it is customary to correct the aperture
velocity dispersions to some fiducial radius, such as the effective
radius ($\Re$), or one eighth of the effective radius. However, doing
so requires making an assumption for the variation of $\sigma$ with
radius. Since we will directly use the strength of the correlation
between $\sigma$ and $\Re$ at fixed mass to constrain our models, we
do not wish to introduce an artificial correlation. Instead, our
approach is to measure observed scaling relations using the
uncorrected fiber velocity dispersions, $\sigmaap$, and to explicitly
compute the velocity dispersion within the SDSS fiber in our models
using the spherical Jeans equations (see Appendix A for details).

We use stellar masses from the MPA/JHU database\footnote{ Available at
  http://www.mpa-garching.mpg.de/SDSS/DR7/}. These are derived by
fitting $ugriz$ photometry with Bruzual \& Charlot (2003) SPS models
assuming a Chabrier (2003) IMF. To account for a non-universal IMF, or
systematic errors in stellar mass measurements, we allow for an offset
between the true stellar masses, $\Mstar$, and these SPS stellar
masses, $\Msps$, which we denote $\Delta_{\rm IMF} \equiv
\log(\Mstar/\Msps)$.

The uncertainty on the stellar masses are nominally $\simeq 0.1$ dex,
but there will be some galaxies with much larger uncertainties.  At
the highest masses, and largest size-offsets, these outliers can bias
the slopes of the scaling relations. As in Dutton, Mendel, \& Simard
(2012) we clean the sample by using the relation between stellar
mass-to-light ratio and velocity dispersion: $\Msps/L_r$ vs
$\sigmaap$. We remove $\sim 700$ galaxies (0.5\% of our early-type
galaxy sample) that are more than $\pm 4\sigma$ from the median
relation. This number is $\sim 100$ times higher than the expected
number of high-$\sigma$ offsets for log-normal scatter in $\Msps/L_r$.

\subsection{Mean Relations}
The observed velocity dispersion - stellar mass (VM) and size -
stellar mass (RM) relations for our sample of SDSS early-type galaxies
are shown in Fig.~\ref{fig:vmr}.  The upper panels shows the median
(filled circles) together with the 16th and 84th percentiles (dotted
lines) of velocity and size in bins of stellar mass (these data points
are given in Table~\ref{tab:datavm}).  The median relations are well
fitted with double power-laws (in the variables $10^x$ and $10^y$):
\begin{equation}
  y= y_0 + \alpha (x-x_0) +\frac{(\beta-\alpha)}{\gamma} \log \left[\frac{1}{2}+\frac{1}{2}10^{\gamma(x-x_0)}\right].
\label{eq:power2}
\end{equation}
Here $\alpha$ is the slope at $x\ll x_0$, $\beta$ is the slope at
$x\gg x_0$, $x_0$ is the transition scale, $y_0=y(x_0)$, and $\gamma$
controls the sharpness of the transition (higher $\gamma$ results in a
sharper transition).  The parameters of the fits are given in
Table~\ref{tab:fits}.  For the velocity-mass relation the slope varies
from $\sim 0.4$ at low masses to $\sim 0.2$ at high masses, while for
the size-mass relation the slope varies from $\sim 0.0$ at low masses to
$\sim 0.7$ at high masses.

The lower panels show the scatter about the median relations (open
circles). For both velocity-mass and size-mass relations the scatter
decreases from low to high mass.  The mass dependence of the scatter
is fitted with the following function:
\begin{equation}
\label{eq:s2}
y = y_2 + \frac{ y_1 - y_2}{1 + 10^{\gamma(x-x_0)}}.
\end{equation}
Here $y_1$ is the asymptotic value at $x \ll x_0$, $y_2$ is the
asymptotic value at $x \gg x_0$, $x_{0}$ is the transition scale, and
$\gamma$ controls the sharpness of the transition.  The values of the
best fit parameters are given in Table~\ref{tab:fits}.

\begin{figure}
\centerline{
\psfig{figure=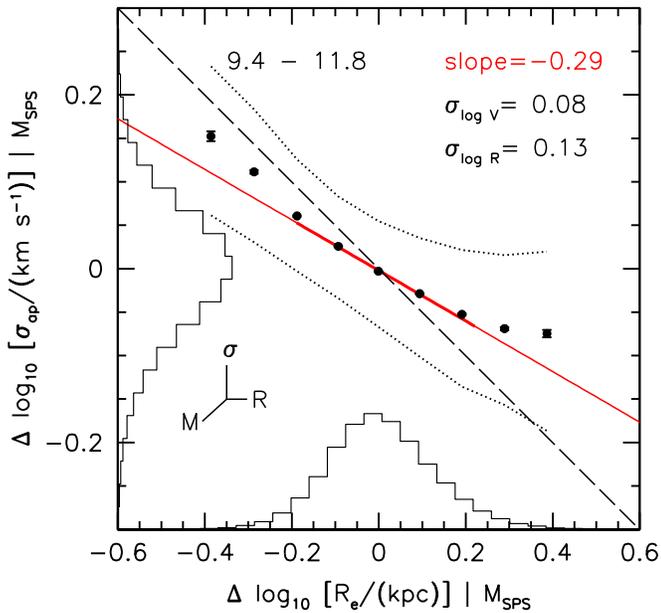,width=0.49\textwidth}
}
\caption{Correlation between residuals of the velocity-mass and
  size-mass relations (Fig.~\ref{fig:vmr}) for all early-type
  galaxies.  The solid line shows the best fit linear relation which
  has a slope of $\dvr=-0.29$. The thick portion of this line shows
  the region fitted, which corresponds to the 5th and 95th percentiles
  of the distribution of size offsets.  The long-dashed line is the
  virial relation with slope $=-0.5$. The compass shows the direction
  the median errors on velocity ($\sigmaap$), size ($\Re$), and mass
  ($\Msps$) scatters galaxies in this plane. This shows that most of
  the scatter about the median relation is due to observational
  errors.}
\label{fig:dvr_all}
\end{figure}

\begin{figure}
\centerline{
\psfig{figure=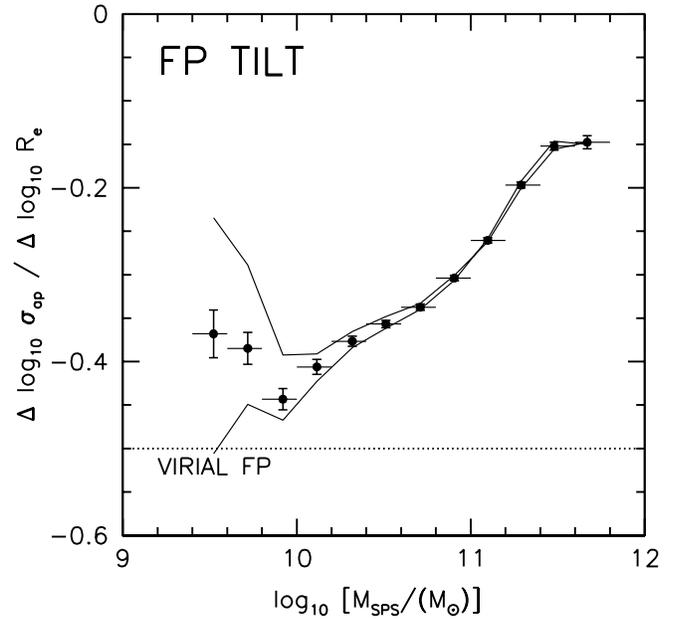,width=0.49\textwidth}
}
\caption{Slopes of the correlation between residuals of the
  velocity-mass ($\Delta\log\sigmaap$) and size-mass ($\Delta\log\Re$)
  relations, as a function of stellar mass. The horizontal bar shows
  the width of the bin in stellar mass, while the vertical error bar
  shows the uncertainty on the slope. The two solid lines show the
  results obtained using the two different set of stellar velocity
  dispersions, indicating systematic errors at low masses.  The virial
  FP has a slope of $-0.5$, and thus we find that the tilt of the
  observed FP is mass dependent.}
\label{fig:dvr_mass}
\end{figure}

\subsection{Residual Relations}

The correlation between the residuals of the VM and RM relations,
$\Delta\log\Re$ and $\Delta\log\sigmaap$, for the full sample of
early-type galaxies is shown in Fig.~\ref{fig:dvr_all}. The histograms
show the distribution of $\Delta\log \Re$ and $\Delta \log
\sigmaap$. The standard deviation of the velocity and size residuals
is 0.08 dex and 0.13 dex, respectively.  The filled circles show the
median $\Delta\log \sigmaap$ in bins of $\Delta\log \Re$, while the
dotted lines show the 16th and 84th percentiles of the distribution.
The median errors on velocity, size, and mass are indicated with the
compass, and shows that most of the scatter about the median relation
is due to observational errors. Note that due to the curvature in the
size-mass and velocity-mass relations, the direction of the stellar
mass error changes from almost vertical at low masses, to almost
horizontal at high masses.

A power-law fit to these data (over the region indicated by the thick
line) results in a slope of $\dvr=-0.29\pm 0.01$. The virial
fundamental plane has $\dvr=-0.5$, which is shown with the diagonal
dashed line. Thus the fact that we observed $\dvr \ne -0.5$ is
equivalent to the statement that the {\it observed} fundamental plane
is tilted with respect to the virial fundamental plane.  The slope
varies with $\Delta \log \Re$, indicating that the tilt of the {\it
  observed} fundamental plane is not a constant.  For large negative
size offsets, $\dvr \simeq -0.5$, suggesting that baryons dominate
within the effective radius, whereas for large positive size offsets,
$\dvr\simeq 0$, suggesting that there is significant dark matter
(Courteau \& Rix 1999). These nominal trends of dark matter fraction
with size offset are qualitatively consistent with the expectations
for galaxies embedded in extended dark matter haloes --- for a fixed
halo mass and stellar mass, smaller galaxies will have lower dark
matter fractions within an effective radius, and hence smaller (more
negative) $\dvr$.

Fig.~\ref{fig:dvr_mass} shows $\dvr$ computed in stellar mass bins of
width 0.2 dex (these data points are given in
Table~\ref{tab:datavm}). This shows that while $\dvr$ is always
negative, it is not a constant, i.e., the tilt of the {\it observed}
fundamental plane increases with increasing stellar mass. This echos
previous studies which show there is curvature to the fundamental
plane (e.g., Zaritsky \etal 2006, 2011; Hyde \& Bernardi 2009;
Tollerud \etal 2011). As an estimate of systematic uncertainties we
measure $\dvr$ using the two sets of velocity dispersions available
for SDSS galaxies. The two measurements are shown with the solid lines
in Fig.~\ref{fig:dvr_mass}. They are in good agreement above
$\Msps\sim 10^{10}\Msun$, but below this mass there are significant
differences.

As a final note we stress that the data shown in
Figs.~\ref{fig:dvr_all} \& \ref{fig:dvr_mass} are not corrected for
aperture effects or measurement errors, which both tend to weaken the
observed correlations. Aperture effects are more important when the
fiber only covers a small fraction of the galaxy light (i.e.,
intrinsically larger galaxies or galaxies at lower
redshifts). Measurement errors in stellar masses are most important
(especially at high masses) as these couple offsets from the size-mass
and velocity mass relations.  Rather than try to correct the data
(which in the case of aperture corrections is model dependent, and
thus non unique), our approach is to explicitly include these effects
in our models.  We will show that an observed $\dvr\sim -0.3$ can be
explained by models in which mass follows light (and thus follow the
viral fundamental plane between $\sigma_{\rm e}$, $\Re$, and
$\Mstar$).

\begin{figure}
\centerline{
\psfig{figure=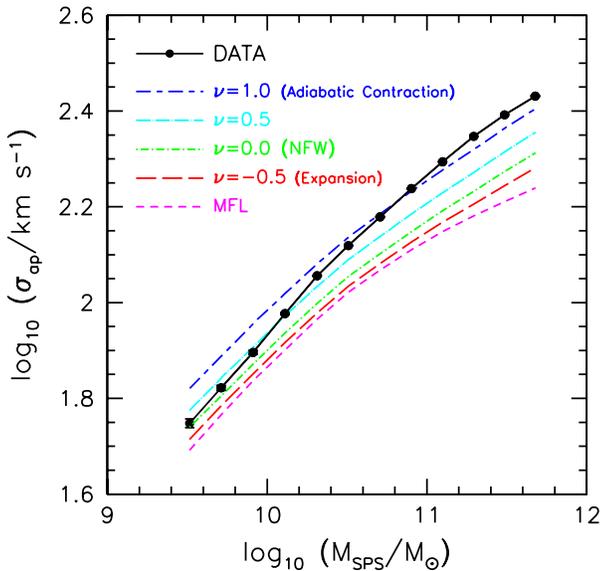,width=0.45\textwidth}
}
\caption{Comparison between observed and model VM relations assuming a
  Chabrier IMF. The filled circles and solid black line shows the
  observed median relation, with error bars indicating the error on
  the median.  The colored lines show the model relations assuming
  isotropic stellar velocity dispersions ($\beta=0$). The models only
  differ in the structure of the dark matter halo, which ranges from
  adiabatically contracted NFW dark matter haloes ($\nu=1$, blue line)
  to no dark matter (MFL, magenta line). None of these models
  reproduces the slope of the VM relation, indicating the need for a
  non-universal IMF and/or non-universal dark halo response.}
\label{fig:vm_chabrier}
\end{figure}

\section{CONSTRAINTS FROM THE VELOCITY-MASS RELATION}

In this section we use the velocity-stellar mass relation to constrain
the two free parameters of our model (see Dutton \etal 2011a and
Appendix A for more details): the stellar mass normalization
$\DeltaIMF=\log(\Mstar/\Msps)$ and dark halo response $\nu$.

\subsection{Universal IMF and universal halo response}

We start by constructing model samples with a universal Chabrier IMF.
We consider four halo responses: standard adiabatic contraction
$\nu=1$ (Blumenthal \etal 1986), reduced halo contraction $\nu=0.5$
(c.f., Abadi \etal 2010); no halo contraction $\nu=0$ (i.e., NFW
haloes); and halo expansion $\nu=-0.5$. In addition we consider a
model in which mass-follows-light (MFL).  For each model we compute
aperture velocity dispersions for 5000 model galaxies evenly spaced in
$\log(\Msps/\Msun)$ (from 9.3 to 11.9), and including log-normal
scatter in sizes, dark halo masses, and dark halo concentrations. We
then re-sample these galaxies (100 times) according to the observed
intrinsic distribution of stellar masses, and add measurement errors
in stellar mass, size and velocity dispersion (see Appendix A). This
procedure results in a sample of $\sim 150\,000$ model galaxies with the
same distribution of stellar masses and sizes as our observed sample.

The median velocity-mass relations of these models are shown in
Fig.~\ref{fig:vm_chabrier}. None of our models is able to reproduce
the observed VM relation, even allowing for zero point offsets
(corresponding to different, but still universal IMFs). There are two
primary solutions to this problem: 1) Allow the stellar mass to vary
from that predicted by a universal Chabrier IMF, with ``heavier'' IMFs
in more massive (i.e., higher $\Msps$) galaxies; or 2) Allow the halo
response to vary with galaxy mass, with e.g., no halo contraction in
low mass galaxies and stronger halo contraction in progressively
higher mass galaxies.

In principle, another solution would be to allow the stellar
anisotropy to vary with galaxy mass, however, observations find no
evidence for a mass dependence to the stellar anisotropy (e.g.,
Gerhard \etal 2001). Furthermore, as we show below, the maximum range
of reasonable anisotropy only has a $\simeq 10\%$ effect on the
derived masses.

We now construct models with non-universal IMFs and/or non-universal
halo responses that reproduce the median VM relation. In
\S~\ref{sec:results} we will use the tilt of the fundamental plane
(Fig.~\ref{fig:dvr_mass}) to distinguish between these models.

\begin{figure*}
  \centerline{\psfig{figure=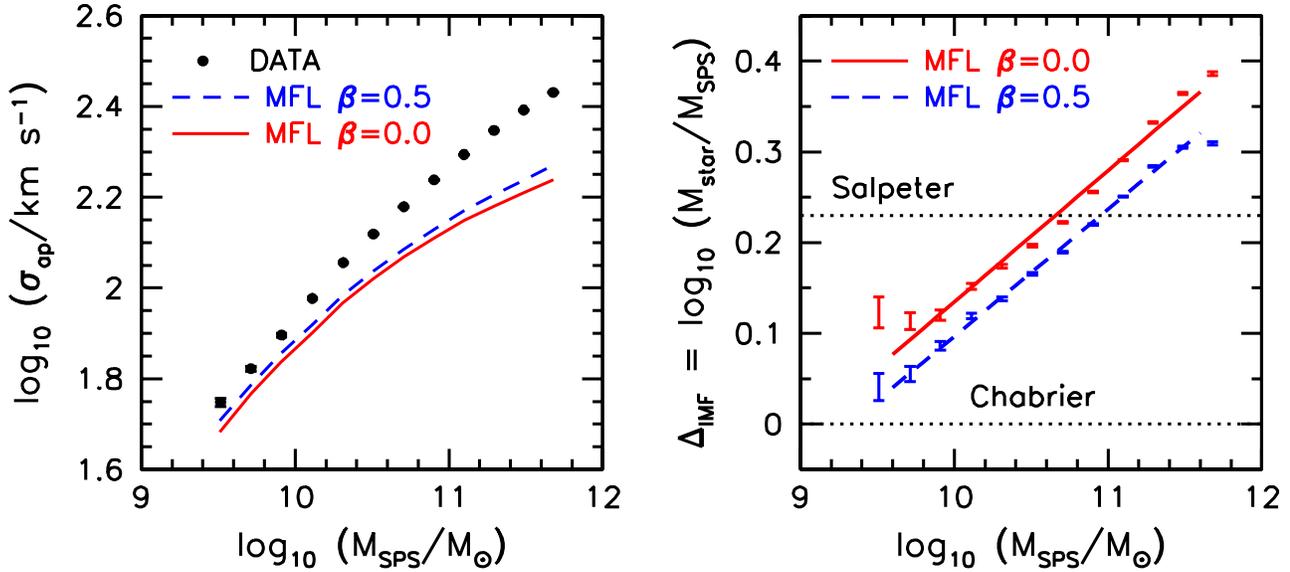,width=0.99\textwidth} }
  \caption{{\it Left:} VM relations for data (black filled circles)
    and mass-follows-light (MFL) models (red and blue lines) assuming
    a Chabrier (2003) IMF. The model galaxies have constant velocity
    anisotropy with $\beta=0$ (red solid line) and $\beta=0.5$ (blue
    dashed line). {\it Right:} Relation between stellar mass
    ($\Mstar$) required for the model to match the observed VM
    relation and stellar population synthesis mass ($\Msps$). The
    offset between these two masses is well fitted with a power-law,
    with parameters as indicated. Interpreting this offset in terms of
    the IMF requires ``heavier'' IMFs in more massive galaxies.}
\label{fig:vm_model_MFL}
\end{figure*}

\begin{figure*}
\centerline{
\psfig{figure=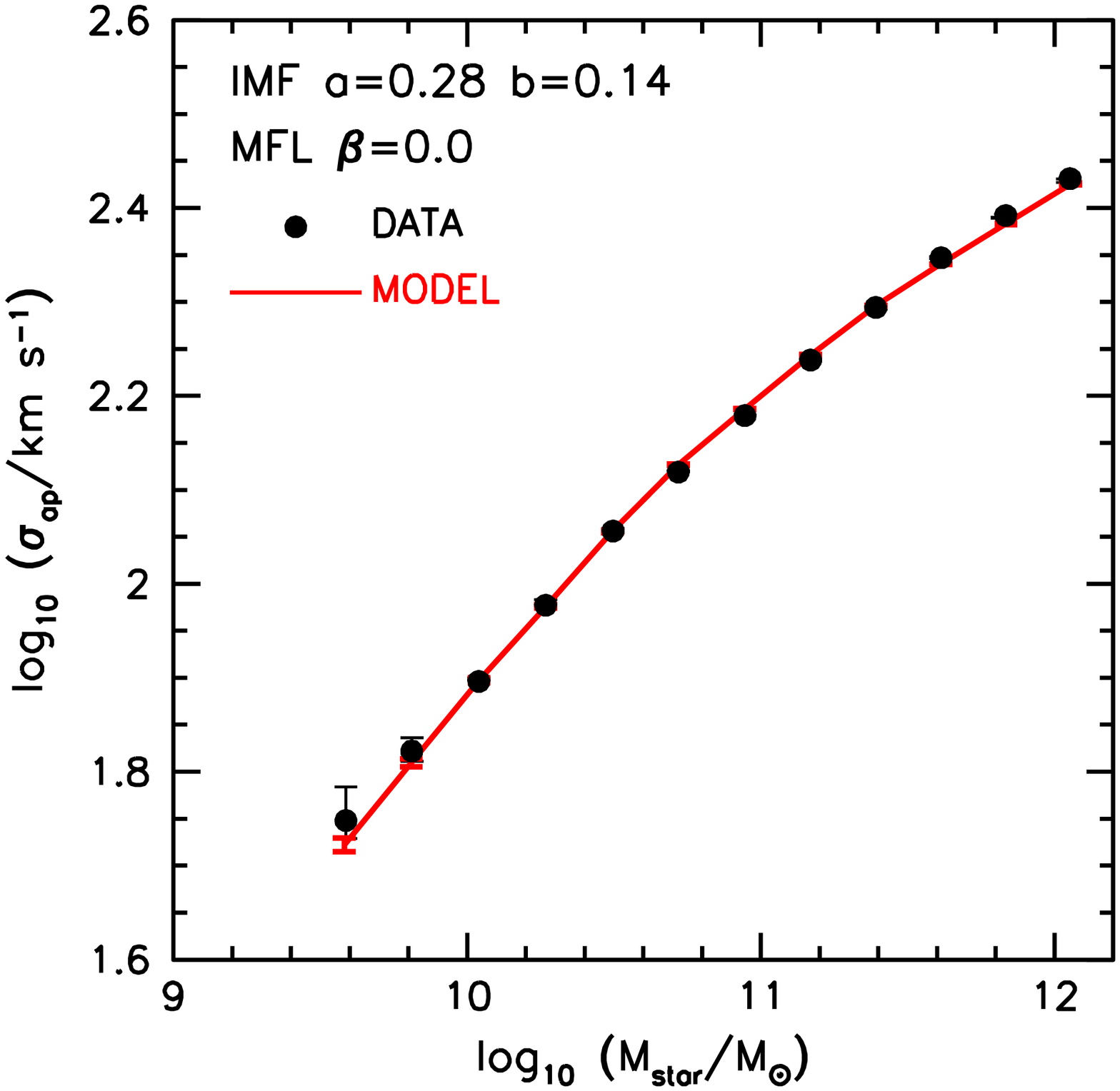,width=0.33\textwidth}
\psfig{figure=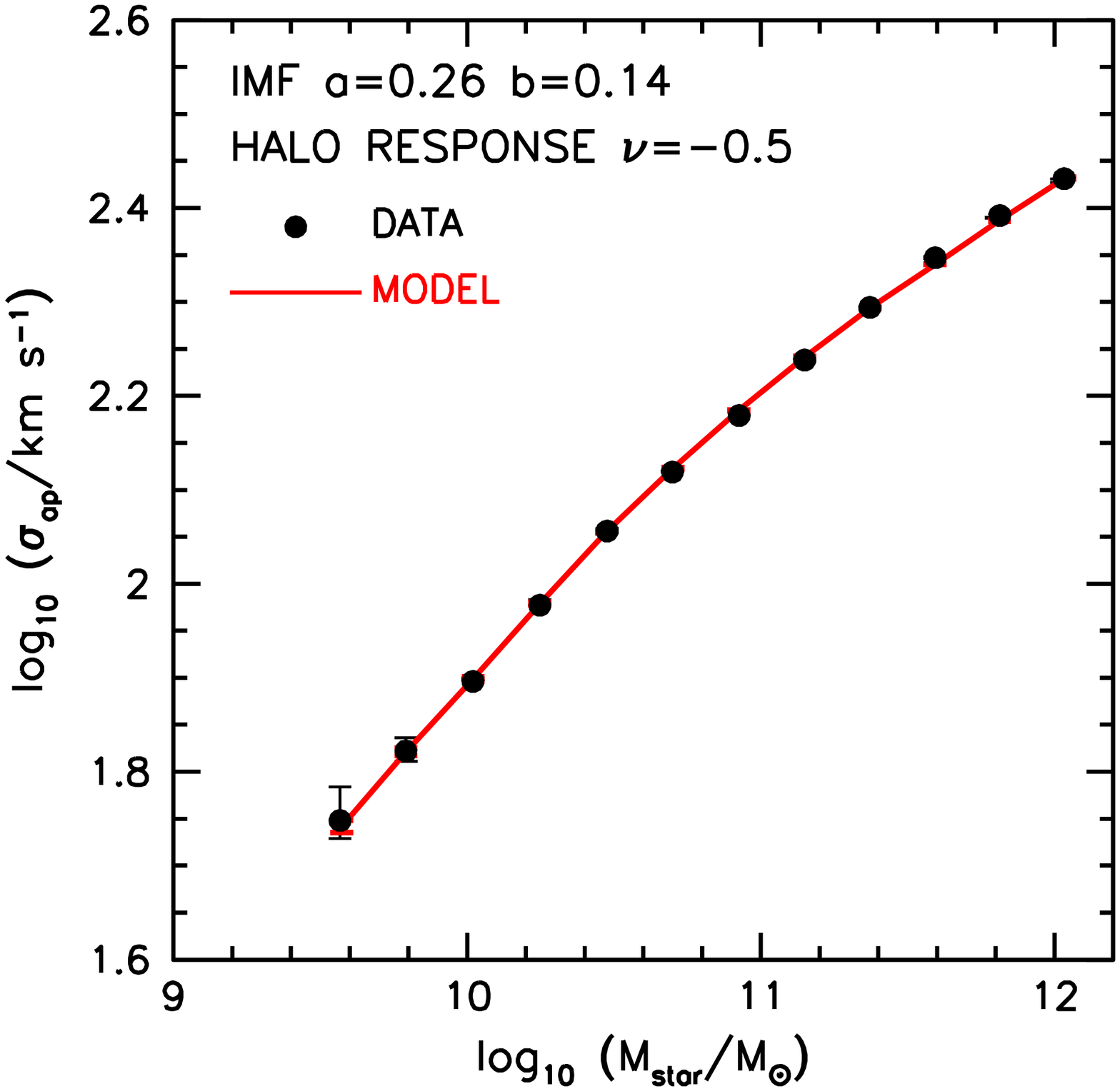,width=0.33\textwidth}
\psfig{figure=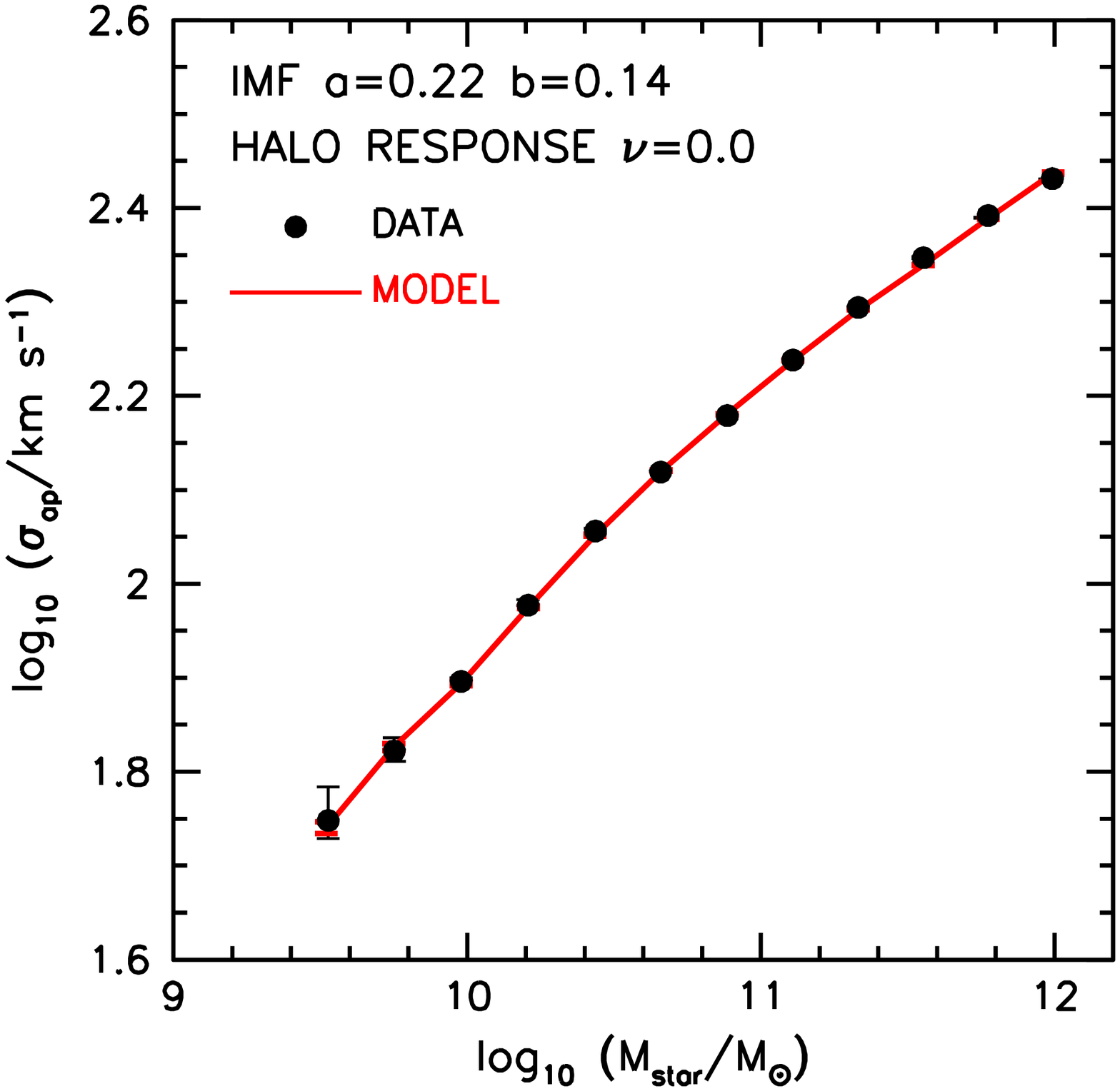,width=0.33\textwidth}
}
\centerline{
\psfig{figure=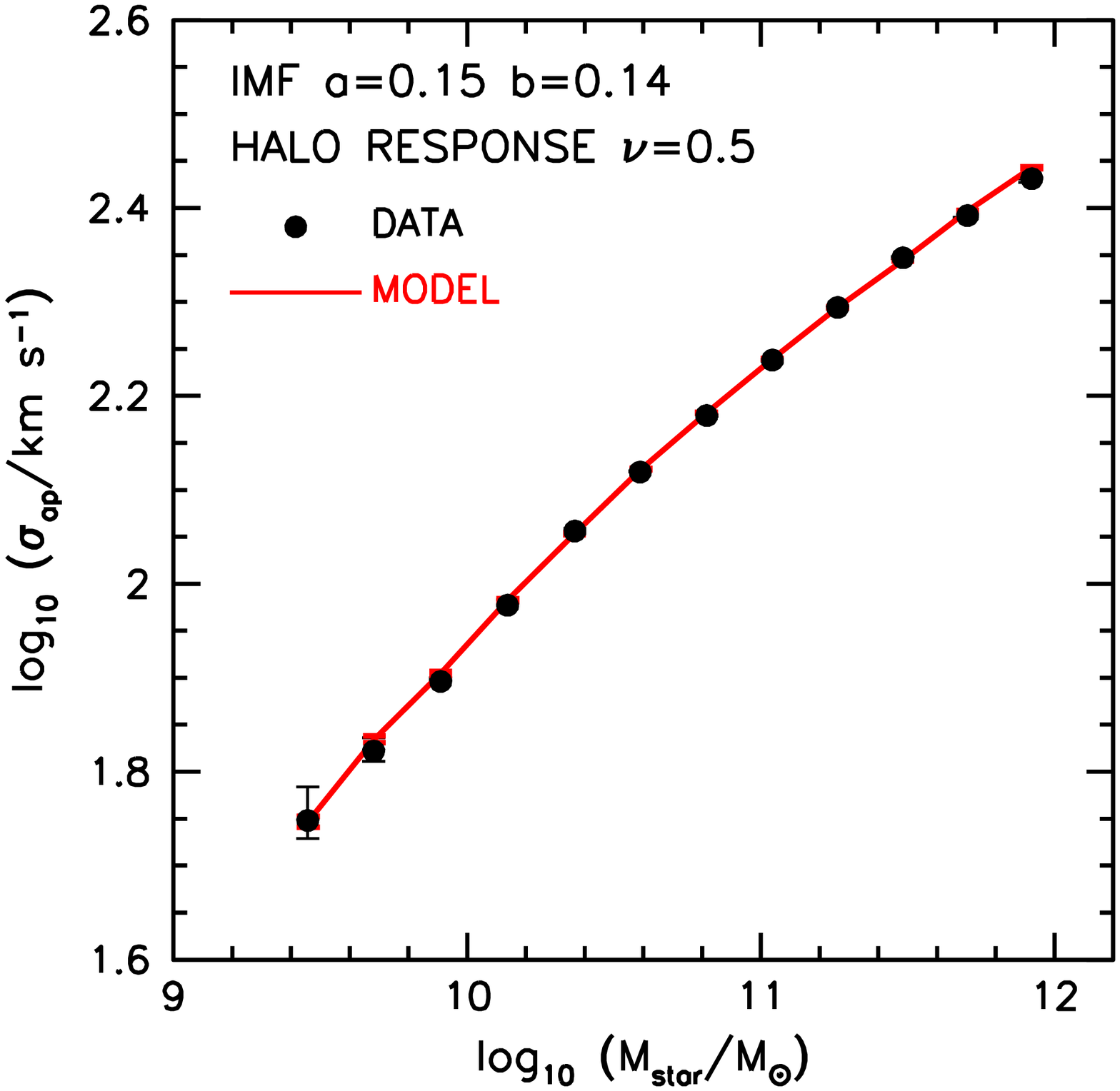,width=0.33\textwidth}
\psfig{figure=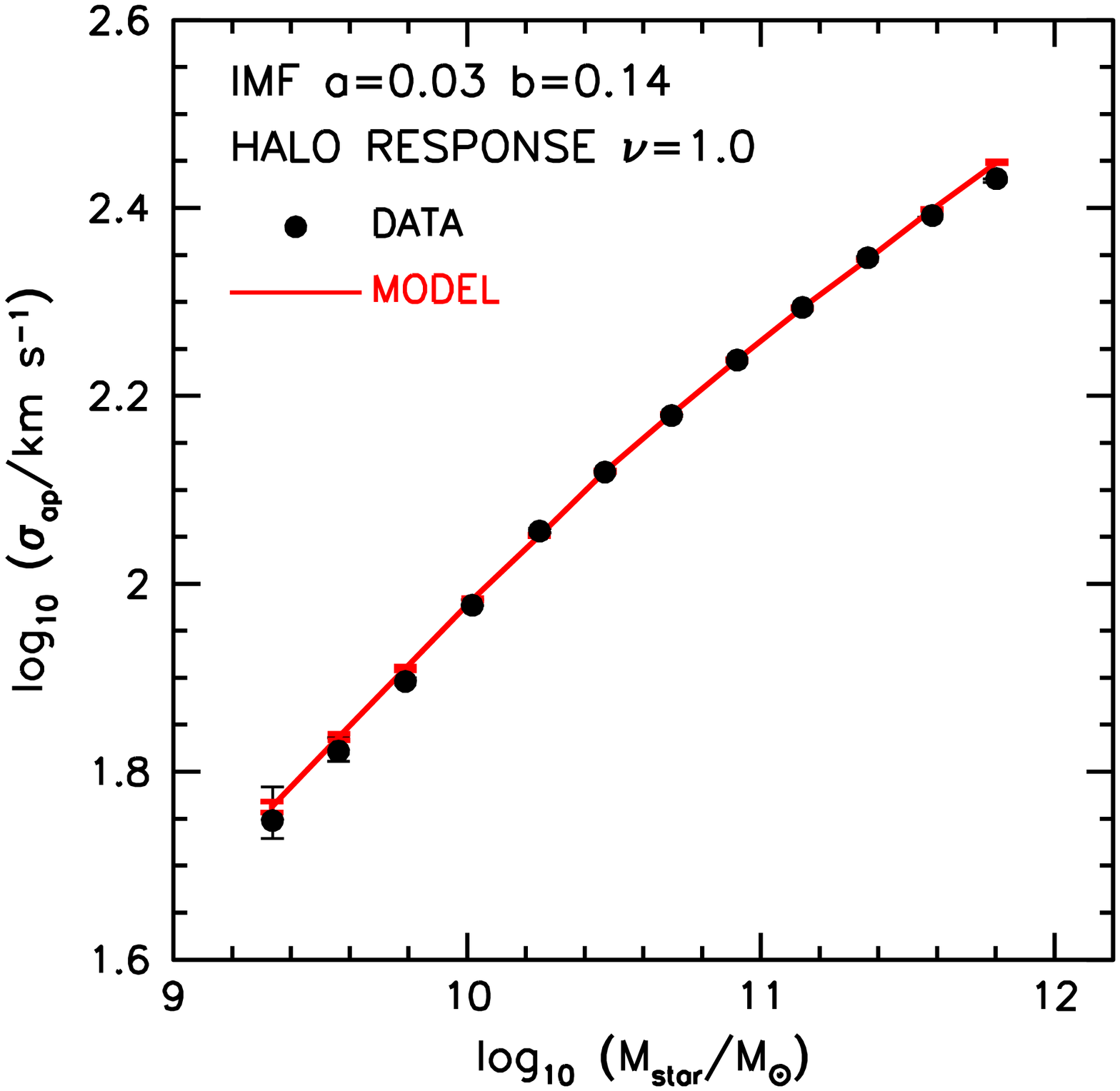,width=0.33\textwidth}
\psfig{figure=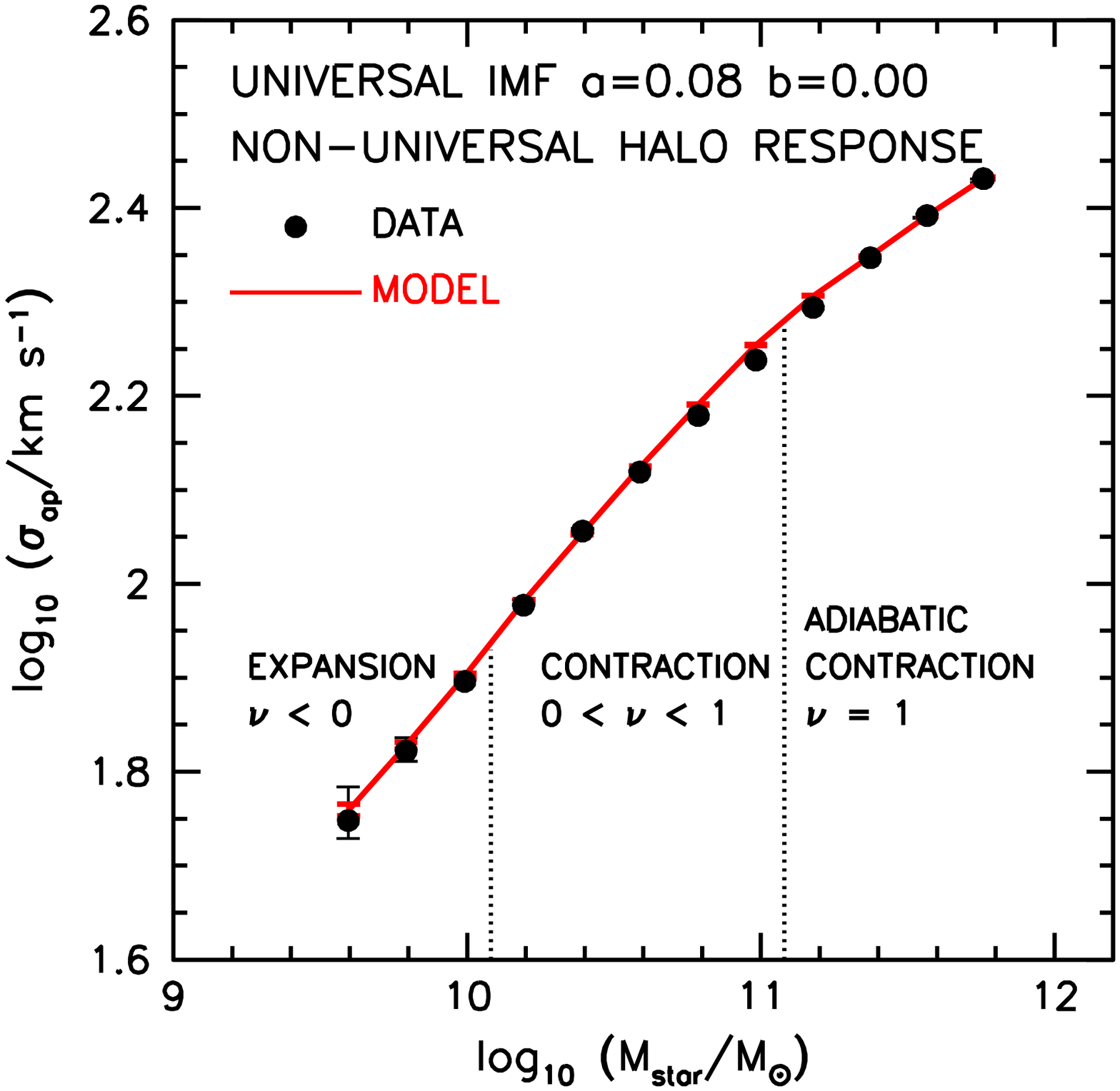,width=0.33\textwidth}
}
\caption{Comparison between observed (black circles) and model (red
  lines) VM relations. Each model has a different halo response
  (parametrized by $\nu$) and stellar mass normalization (parametrized
  by $a$ and $b$). Models with stronger halo contraction require
  lighter IMFs (lower $a$). For a universal halo response (i.e., fixed
  $\nu$) a non-universal IMF ($b\ne 0$) is required. }
\label{fig:vm_noscatter}
\end{figure*}

\subsection{Non-universal IMF with mass-follows-light}
For MFL models we can easily calculate the stellar masses required to
match the observed velocity dispersions using
\begin{equation}
  \Mstar=\Msps(\sigmaap/\sigma_{\rm ap,SPS})^2,
\end{equation}
where $\sigmaap$ is the observed velocity dispersion and $\sigma_{\rm
  ap,SPS}$ is the model velocity dispersion computed assuming $\Msps$.
The stellar mass offset, $\Delta_{\rm IMF}$, as a function of SPS mass
for two MFL models (with different anisotropy) is shown in the right
panels of Fig.~\ref{fig:vm_model_MFL}.  For $10^{9.7} \lta \Msps \lta
10^{11.5}$ the stellar mass offset is well fitted with a linear
relation:
\begin{equation}
\Delta_{\rm IMF} = \log(\Mstar/M_{\rm SPS}) = a + b\log(\Msps/10^{11}\Msun).
\end{equation}
For both isotropic ($\beta=0$) and radially anisotropic orbits (with
$\beta=0.5$) we find a slope of $b\simeq0.14\pm0.01$. This results in a
factor of $\simeq 2$ difference in $\Delta_{\rm IMF}$ across the range
of $\Msps$ that we study. In terms of normalization, for
$\beta=(0.0,0.5)$ we find $a=(0.278,0.238)$, which is slightly higher
than a Salpeter IMF ($\Delta_{\rm IMF}\simeq 0.23\pm0.01$).

Previous studies have determined the slope, $b$, for early-type
  galaxies under the assumption of MFL, spherical symmetry, and
  isotropic orbits, but using the S\'ersic profile to parametrize the light.
  For example, the result from Trujillo, Burkert \& Bell (2004; T04)
  is equivalent to $b\simeq 0.03\pm0.04$ (assuming a correction of
  0.07 for stellar population effects), while the result from Taylor
  \etal (2010; T10) is equivalent to $b\simeq 0.09\pm0.10$ (where the
  error is dominated by systematics).
  Our result of $b\simeq 0.14\pm0.01$ is consistent with the latter,
  and shows that the choice of how one parametrizes the light profile
  is not critical. While the result from T04 is statistically
  different from ours, they use a very small sample (45 galaxies
  compared to $\sim 1800$ for T10, and $\sim 150\,000$ by us) and thus
  are subject to cosmic variance and/or environmental selection
  effects.

  While it is encouraging that our result is consistent with previous
  studies, it should be stressed that the assumption of
  mass-follows-light is known to be false in massive early-type
  galaxies (e.g., Koopmans \etal 2006, 2009; Gavazzi \etal 2008) --
  below we will show that it is also unable to account for the tilt of
  the fundamental plane in massive early-type galaxies.  Thus this
  exercise in building mass-follows-light models and interpreting the
  tilt of the fundamental plane is somewhat academic (at least for the
  most massive galaxies).
 
  Furthermore, as shown in Appendix~\ref{sec:nonhomology}, there is a
  fundamental inconsistency in the methodology of Taylor \etal
  (2010). Fig.~\ref{fig:nonhomology} shows the effects of non-homology
  on dynamical masses are much weaker when one uses velocity
  dispersions measured within the effective radius, $\sigma_{\rm e}$,
  compared to one eighth an effective radius, $\sigmaei$. However, the
  standard aperture corrections (used by e.g., Taylor \etal 2010)
  result in a constant offset of 0.060 dex between $\sigmae$ and
  $\sigmaei$. Thus if T10 had corrected the SDSS fiber velocity
  dispersions to $\sigmae$ and then applied the same assumptions of
  MFL and isotropy, they would have arrived at different conclusions
  regarding the importance of non-homology in deriving dynamical
  masses.  We do not have this inconsistency in our models because we
  explicitly model the fiber velocity dispersions.

\begin{figure*}
\centerline{ 
\psfig{figure=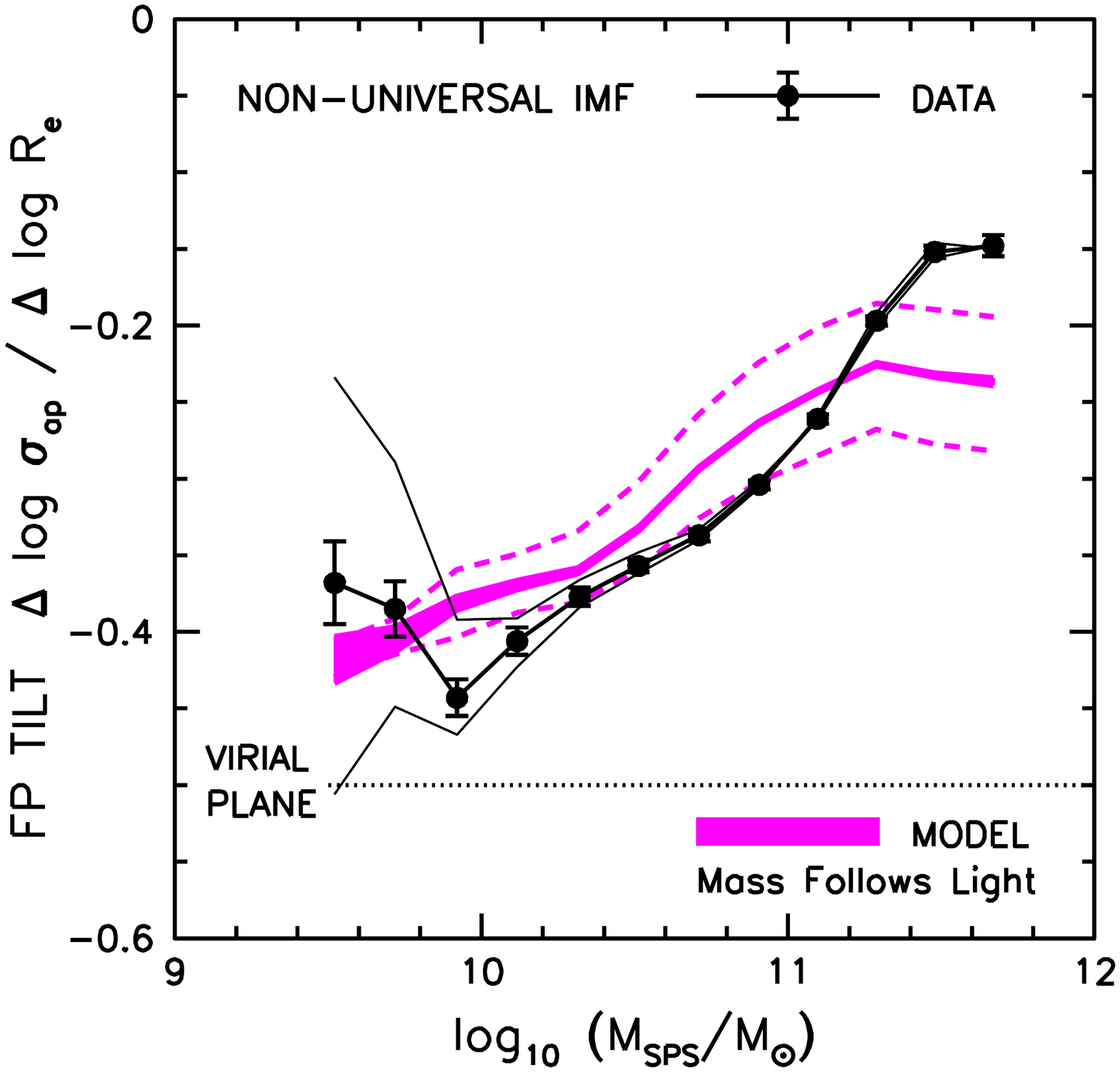,width=0.33\textwidth} 
\psfig{figure=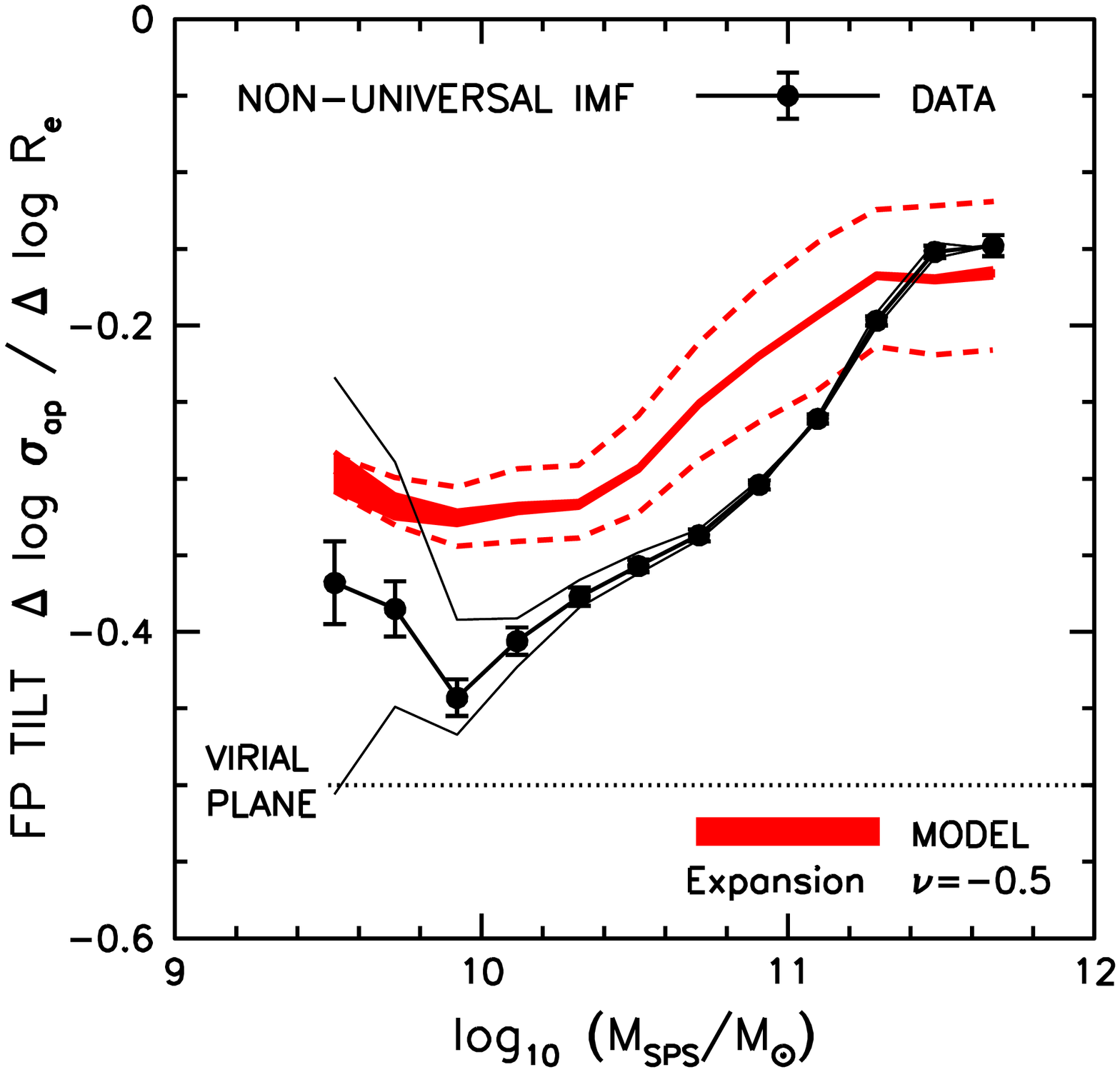,width=0.33\textwidth} 
\psfig{figure=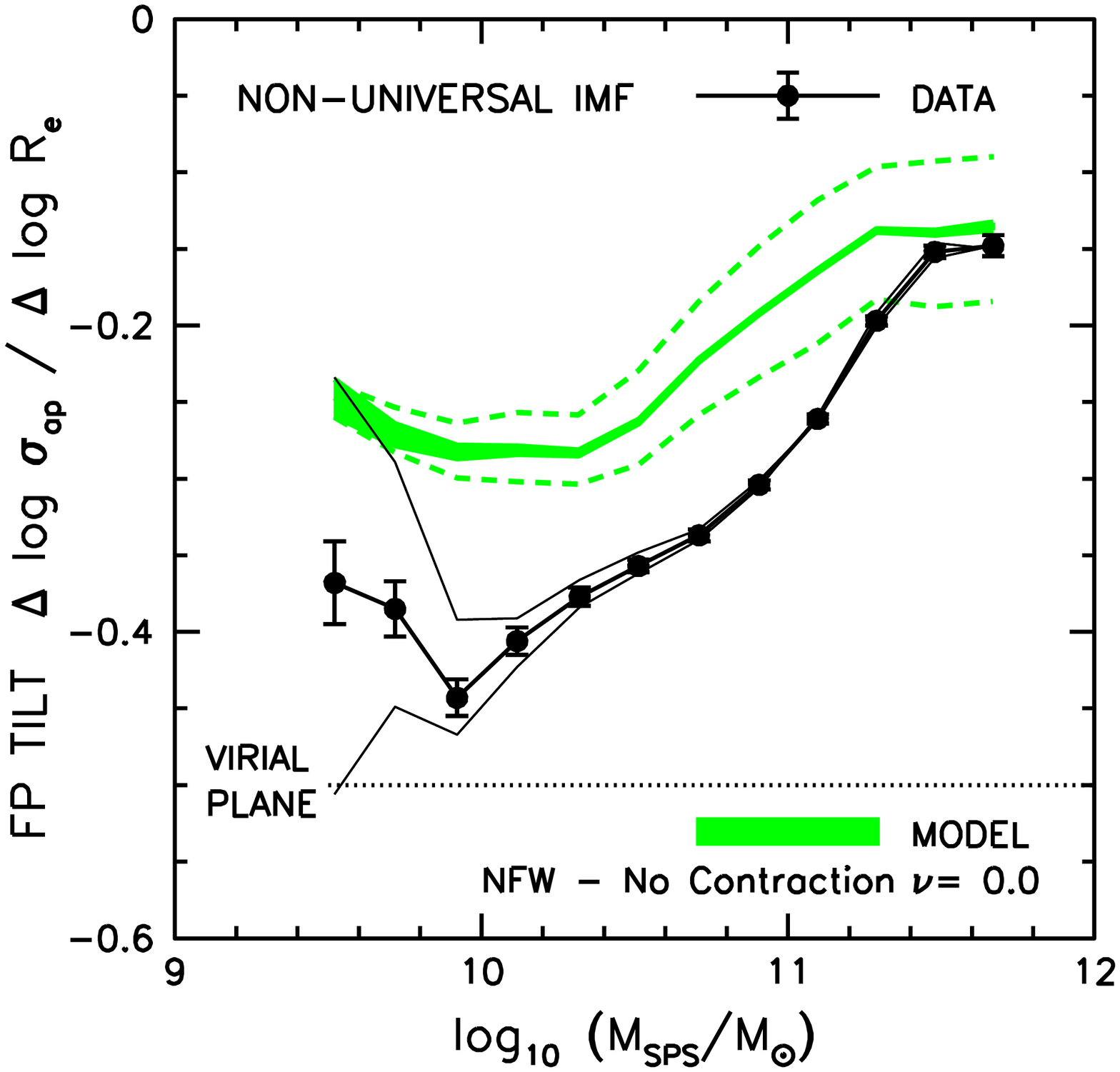,width=0.33\textwidth} 
}
\centerline{
\psfig{figure=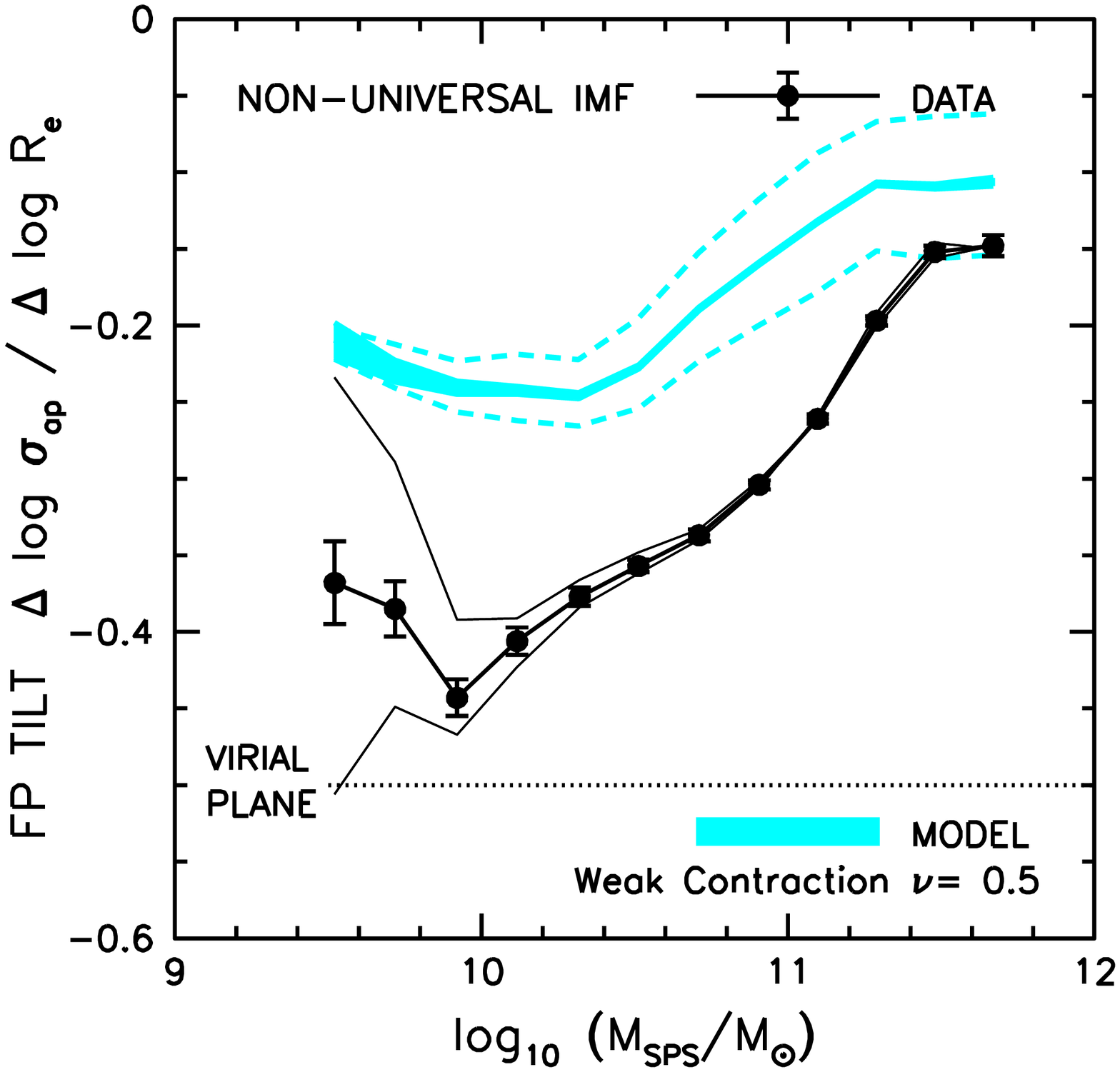,width=0.33\textwidth} 
\psfig{figure=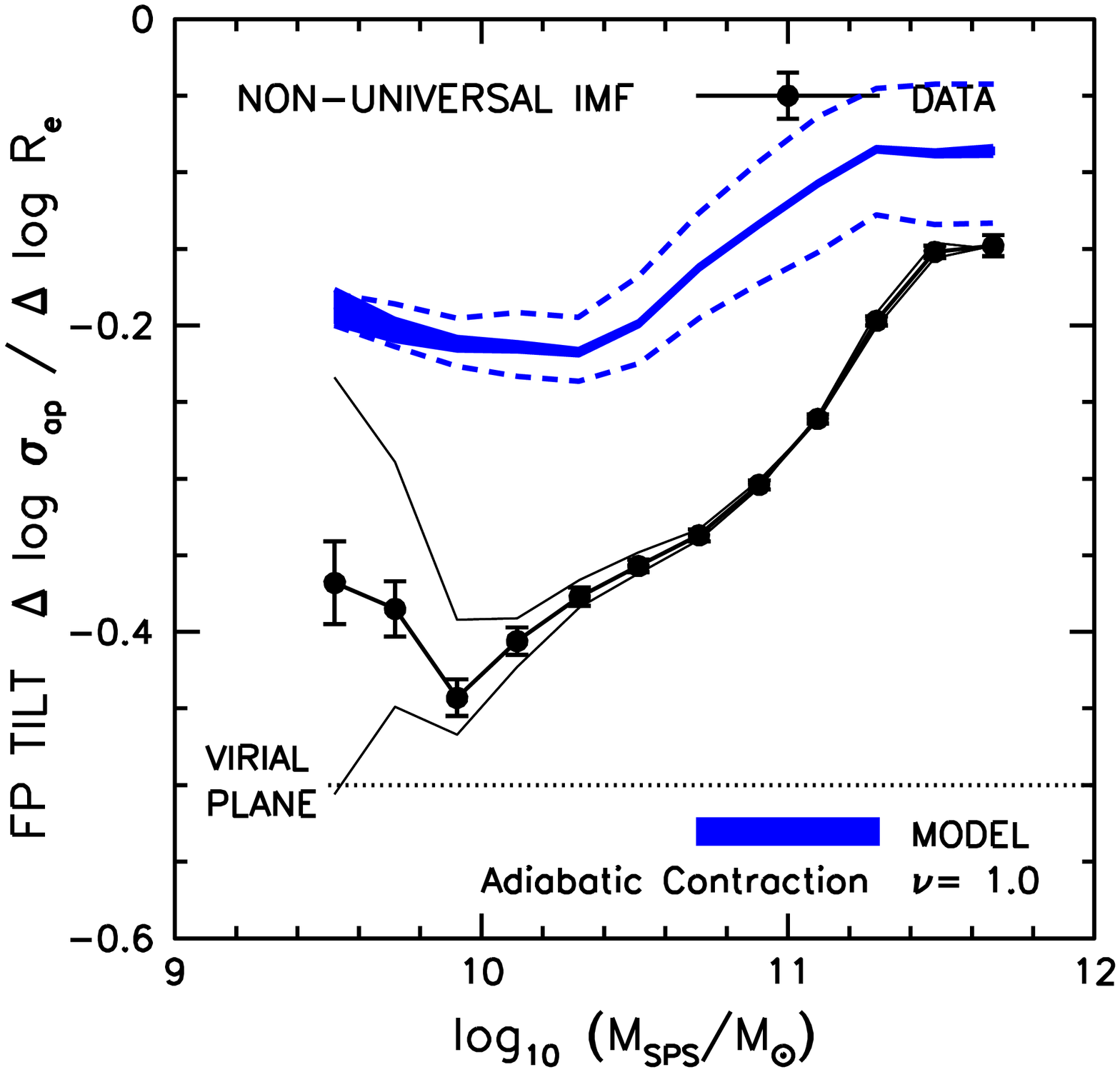,width=0.33\textwidth} 
\psfig{figure=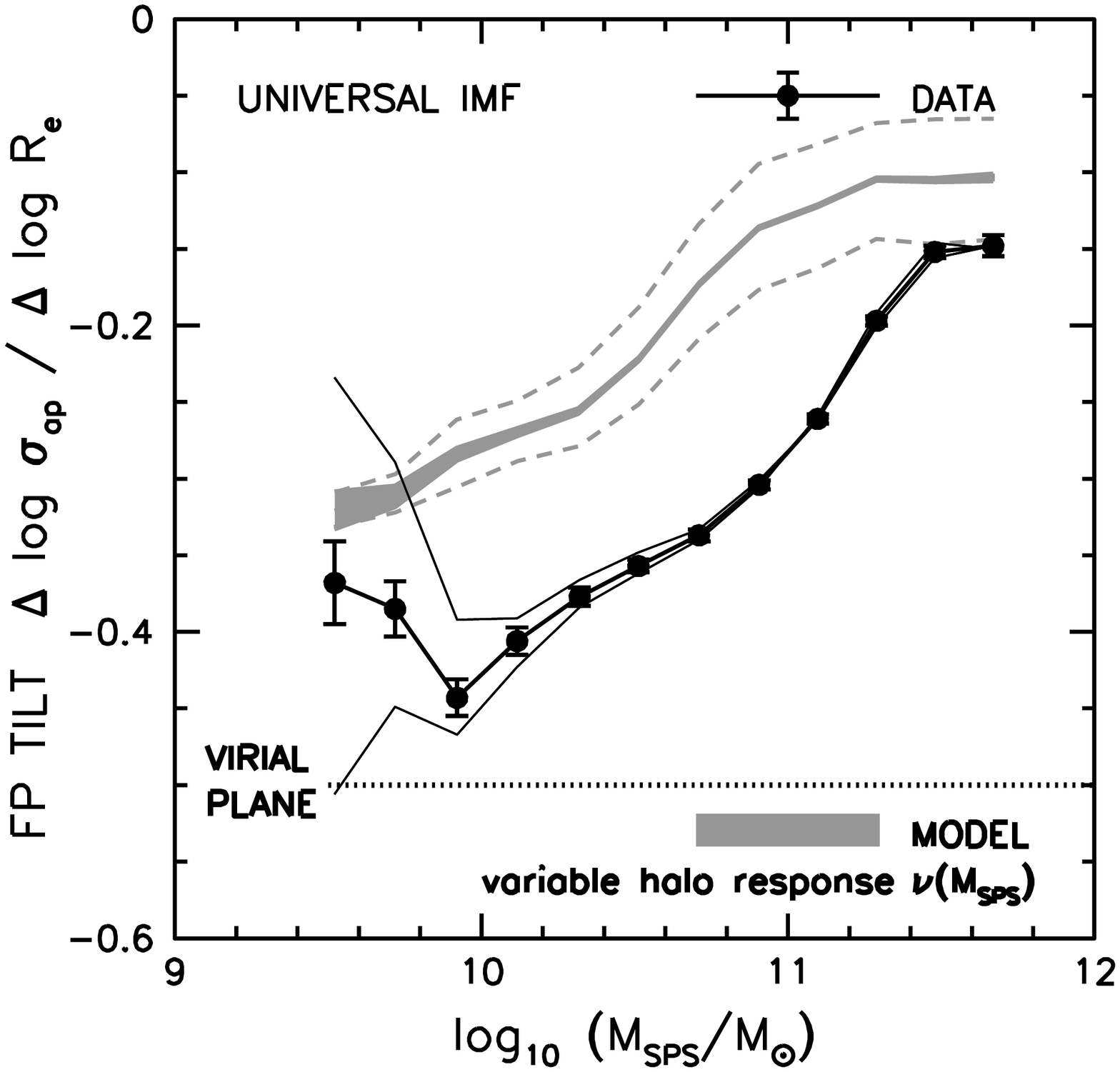,width=0.33\textwidth} 
}
\caption{{Fundamental Plane (FP) tilt vs stellar mass.} Slopes of the
relation between residuals of the VM and RM relations ($\dvr$) vs
stellar mass (normalized to a Chabrier IMF) for models in
Fig.~\ref{fig:vm_noscatter}. The data (black points) is as in
Fig.~\ref{fig:dvr_mass}. For the models the shaded region corresponds
to the fiducial measurement errors, while the dashed region shows 20\%
higher and lower errors on stellar masses.  For all but the highest
masses the mass-follows-light (MFL) model (magenta, upper left panel)
fits the data the best. For the highest masses, models with no halo
contraction are favored, but mild halo contraction is consistent with
the data. Thus reproducing the tilt of the fundamental plane requires
{\it both} and non-universal IMF and non-universal halo response.}
\label{fig:dvmr_model_all}
\end{figure*}

\begin{table}
 \centering
 \caption{Parameters of models, where $\DeltaIMF = a + b \log(\Msps/[10^{11}\Msun])$.}
  \begin{tabular}{cccc}
    \hline
    \hline  
    Halo Response & Anisotropy & $\Delta_{\rm IMF}$ zero & $\Delta_{\rm IMF}$ slope \\
            $\nu$ &  $\beta$    & $a$             & $b$ \\
    \hline
  Variable (see text)  & 0.0 & 0.08 & 0.00\\ 
   1.0  & 0.0 & 0.03 & 0.14\\
   0.5  & 0.0 & 0.15 & 0.14\\
   0.0  & 0.0 & 0.22 & 0.14\\
 $-$0.5 & 0.0 & 0.26 & 0.14\\
    MFL & 0.0 & 0.28 & 0.14\\
    MFL & 0.5 & 0.24 & 0.14\\
\hline
\hline
\label{tab:models}
\end{tabular}
\end{table}

\subsection{Degeneracy between IMF and halo response}

For models with dark matter haloes the model velocity dispersion
depends non-trivially on the stellar mass, and thus when we change
$\Delta_{\rm IMF}$ or the halo response, we need to re-calculate the
model. We determine the IMF offset parameters ($a,b$) for such models
with a grid search.  The stellar mass offset parameters for our suite
of models are given in Table~\ref{tab:models}, and the VM relations
are shown in Fig.~\ref{fig:vm_noscatter}.  All models require a
variable IMF, but the mass dependence, $b=0.14$, is consistent
with being independent of the halo response model. At the reference
stellar mass ($\Msps=10^{11}\Msun$) standard adiabatic contraction
($\nu=1$, Blumenthal \etal 1986) requires an IMF close to Chabrier,
while no dark halo contraction ($\nu=0$) requires an IMF close to
Salpeter.  This is in agreement with our earlier study Dutton \etal
(2011a).

One can construct models that match the VM relation with a universal
IMF, but a non-universal halo response. An example of such a model is
shown in the lower right panel of Fig.~\ref{fig:vm_noscatter}. The IMF
normalization $a=0.08$ is only slightly heavier than a Kroupa (2001)
IMF. The halo response parameter depends on stellar mass as $\nu=1.0
+1.0(\log\Msps -11)$, with a maximum of $\nu=1$. So this model has
adiabatic contraction for $\Msps > 10^{11}\Msun$, and no-contraction
at $\Msps = 10^{10}\Msun$.

The most relevant study in the literature to ours is by Auger \etal
(2010a). These authors fitted \LCDM based models to a sample of $\sim
50$ massive ($\Msps=10^{11.35\pm0.20}\Msun$) elliptical galaxy strong
lenses from the SLACS survey (Bolton \etal 2006). Observational
constraints were masses from strong lensing, aperture stellar velocity
dispersions from SDSS, and weak lensing data from the Hubble Space
Telescope. Auger \etal (2010a) parametrized the relation between
$\Msps$ and $\Mstar$ by $\log \Msps=\log \Mstar - \eta(\log \Mstar-11)
-\alpha$.  In terms of our parametrization: $a=\alpha/(1-\eta)$ and
$b=\eta/(1-\eta)$.  Their best fitting model had uncontracted NFW
haloes and a mass dependent IMF with $\alpha=0.03\pm 0.03$ and
$\eta=0.08\pm 0.04$, where the reference IMF was Salpeter
(1955). Converting to a Chabrier (2003) IMF results in
$a=0.26\pm0.06$, $b=0.09\pm0.06$, which is consistent with our results
for the masses our respective studies overlap.


\section{CONSTRAINTS FROM THE TILT OF THE FUNDAMENTAL PLANE}
\label{sec:results}

We now use the tilt of the fundamental plane to break the degeneracy
between IMF and halo response. Fig.~\ref{fig:dvmr_model_all} shows a
comparison between $\dvr$ and stellar mass for models and
observations.  The observations are shown with black points, lines and
error bars (as in Fig.~\ref{fig:dvr_mass}).  The models are shown with
colored shaded regions, as indicated. The models include measurement
errors in stellar masses, sizes and velocity dispersions. The effect
of measurement errors is to increase the strength of fundamental plane
tilt (i.e., $\dvr$ becomes less negative). Errors on stellar masses
have the strongest effect on $\dvr$. To indicate the dependence of our
results to stellar mass errors the dashed lines show models with
$20\%$ higher and lower measurement errors on stellar masses (i.e.,
varying from 0.08 dex to 0.12 dex).

\begin{figure}
  \centerline{ \psfig{figure=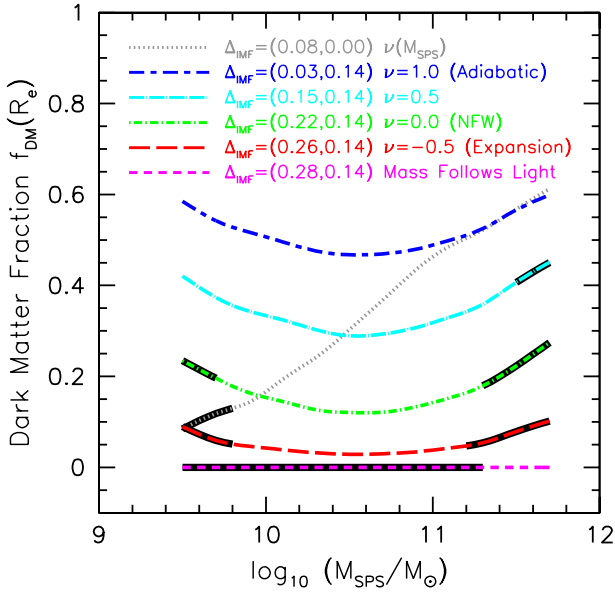,width=0.47\textwidth} }
  \caption{Dark matter fraction within an effective radius vs SPS
    mass. A universal IMF model (dotted grey line) requires dark
    matter fractions increasing with increasing mass, as found by
    numerous previous studies. Universal halo response models (colored
    lines) result in roughly constant dark matter fractions with a
    minimum at $\Msps\sim 3\times 10^{10}\Msun$. The solid black lines
    indicate the models which provide acceptable fits to the
    fundamental plane constraints (see
    Fig.~\ref{fig:dvmr_model_all}).}
\label{fig:dm}
\end{figure}

The first thing that is apparent is that all of the models predict
that $\dvr$ should vary with stellar mass, with (generally) more
negative $\dvr$ at lower masses, in qualitative agreement with the
observations.  However, none of the models reproduces the shape and
normalization of $\dvr$ vs $\Msps$ in detail.  In particular models
with standard adiabatic contraction (blue line, lower middle panel) or
a universal Milky Way type IMF (grey line, lower right panel) are the
most discrepant, under predicting the strength of $\dvr$ by up to
$\sim 0.2$. These models have the ``lightest'' IMF normalizations, and
hence the highest dark matter fractions ($\simeq 50\%$) within an
effective radius (See Fig.~\ref{fig:dm}). The increased dark matter
has the effect of dampening the changes in velocity that are expected
from changing the sizes of the galaxies.

Below a stellar mass of $\Msps \sim 10^{11.2}$ the MFL model (magenta
line, upper left panel) is the only one that is consistent with the
data. At the lowest masses ($\Msps \sim 10^{9.6}$) there is
significant uncertainty in the observations, and while MFL models fit
best, they are not required. Above $\Msps \sim 10^{11.2}$ the MFL
model progressively over-predicts the strength of $\dvr$. Reproducing
the observations thus requires a mass dependent halo response:
expansion (red line) for masses $10^{11.2} \lta \Msps \lta 10^{11.4}$,
and uncontracted NFW haloes (green line) for $\Msps\sim
10^{11.5}$. Given reasonable uncertainties in the stellar mass errors,
models with contracted dark matter haloes (cyan and blue) are
consistent with the data in the most massive galaxies $\Msps \gta
10^{11.4}$.

The fact that MFL does not reproduce the fundamental plane for the
most massive galaxies is consistent with results from strong
gravitational lensing, which find average total mass density slopes
that are close to isothermal $\gamma\simeq -2.08\pm0.03$ (Auger \etal
2010b), and thus shallower than MFL ($\gamma\sim -2.3$ for a Hernquist
profile).

\begin{figure}
  \centerline{ \psfig{figure=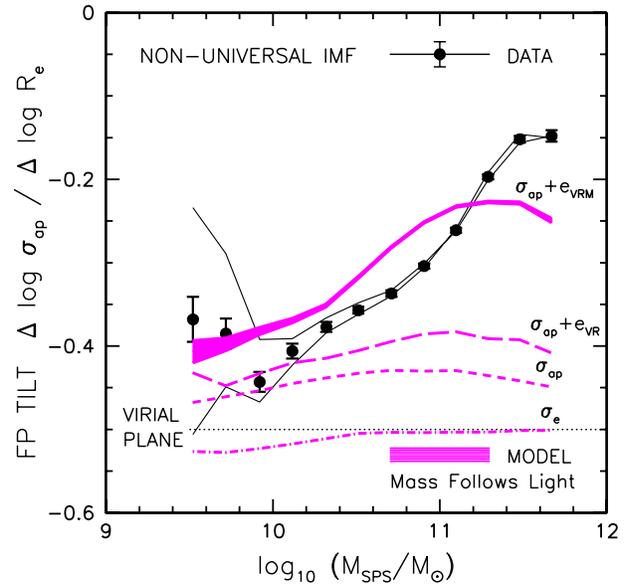,width=0.47\textwidth} }
  \caption{Effect of aperture and measurement errors on the
    fundamental plane tilt for models with mass-follows-light. A model
    with velocity dispersions measured within the effective radius,
    $\sigma_{\rm e}$, closely follows the virial relation (dot-dashed
    line). Using aperture velocity dispersions, $\sigmaap$, results in
    a weaker FP correlation by about $\simeq 0.06$ (short-dashed
    line). Measurement errors in sizes contribute $\simeq 0.04$ to the
    observed tilt (long-dashed line), while measurement errors on
    stellar masses (solid line) contribute up to $\simeq 0.15$ of the
    tilt at the highest masses.}
\label{fig:errors}
\end{figure}

It might seem surprising that MFL models, which follow the virial
fundamental plane, can reproduce the observed fundamental plane
tilt. This can be explained by two effects which are shown for MFL
models in Fig.~\ref{fig:errors}. The first, as mentioned above, is
that measurement errors (on stellar masses and sizes) increase the
observed fundamental plane tilt (i.e., they make $\dvr$ less
negative). In Fig.~\ref{fig:errors} compare the solid and long-dashed
lines for the effect of stellar mass errors, and the short-dashed and
long-dashed lines for the effect of size errors. Note that errors on
velocity dispersions do not contribute to the observed fundamental
plane tilt.  The second is a result of measuring the velocity
dispersions within a fixed physical aperture, rather than within a
relative aperture such as an effective radius. In
Fig.~\ref{fig:errors} compare the dot-dashed line (which uses model
velocity dispersions within the effective radius, $\sigma_{\rm e}$)
with the short-dashed line (which uses model aperture velocity
dispersions, $\sigmaap$).  Each of these effects accounts for roughly
half of the observed tilt of the fundamental plane for intermediate
mass galaxies. The larger tilt at higher masses is mostly due to the
curvature in the size-mass relation which increases the impact of
stellar mass errors.

\subsection{Mass dependence and literature comparison}

\begin{figure*}
 \centerline{ \psfig{figure=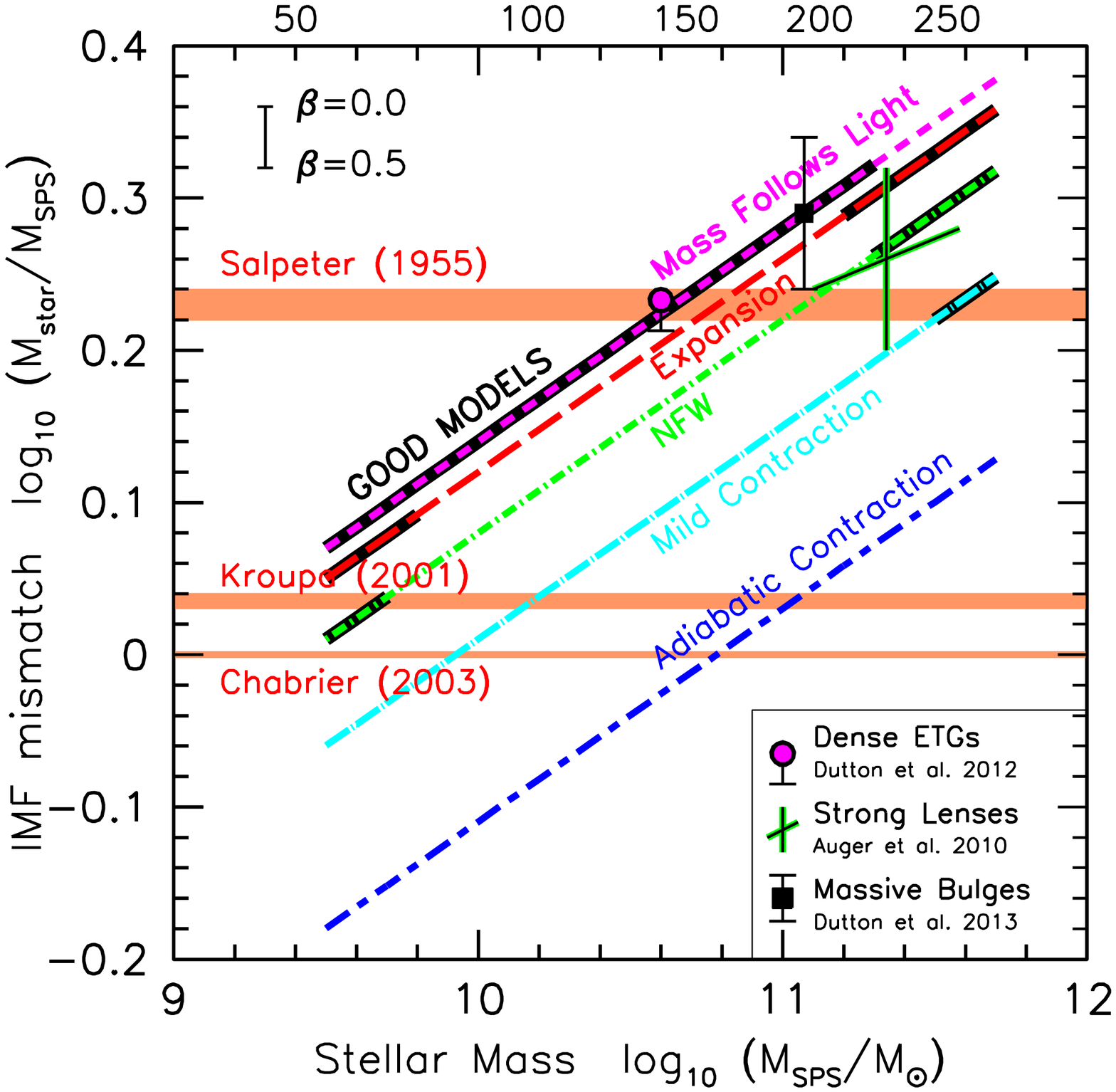,width=0.49\textwidth}
 \psfig{figure=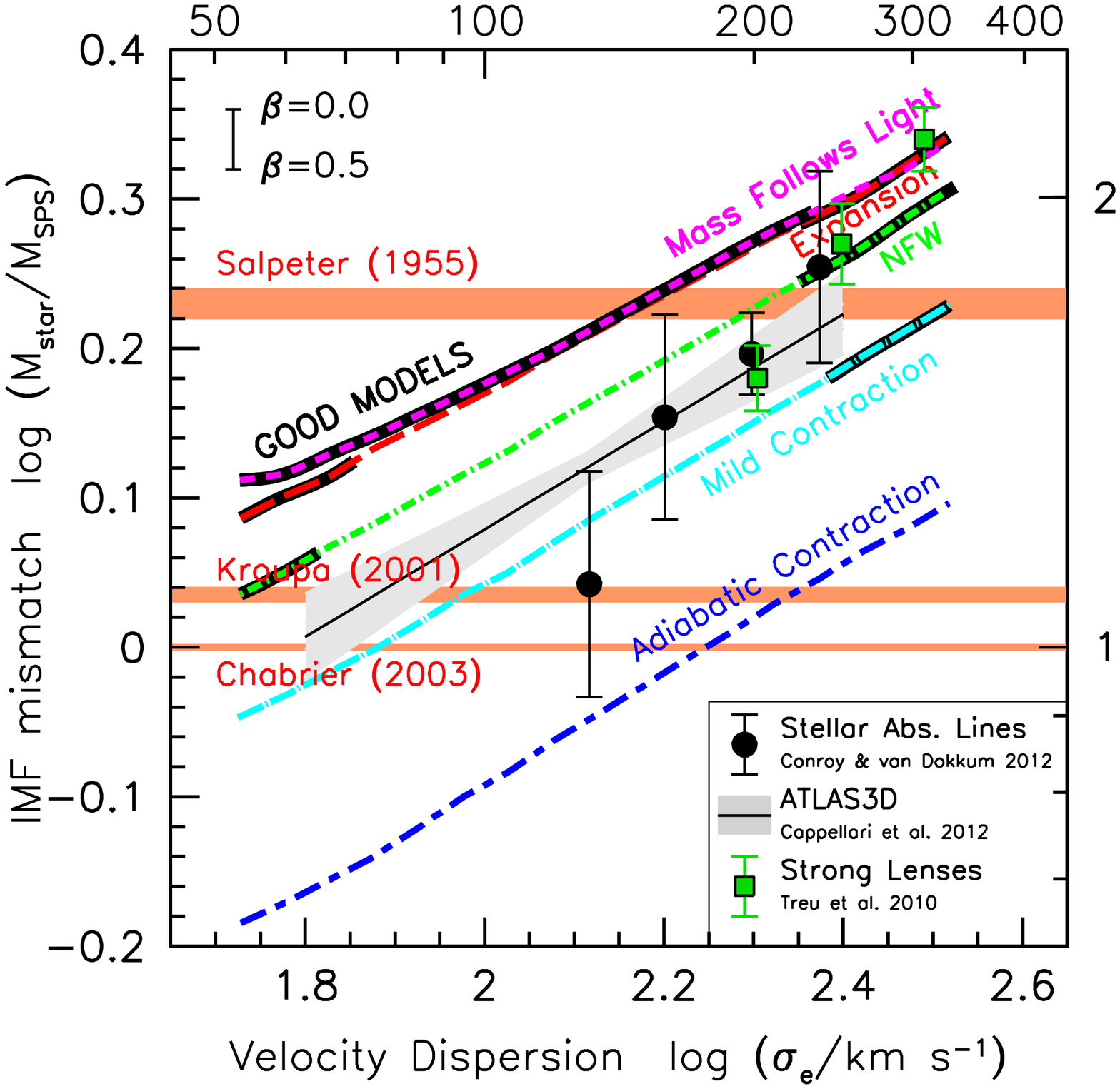,width=0.49\textwidth} }
\caption{Summary of the main results for the stellar IMF and halo
  response of this paper. Reproducing the average Velocity - Mass
  (Faber-Jackson) relation is degenerate between IMF and dark halo
  response. For a universal dark halo response, a non-universal IMF is
  required (colored lines), with $\Mstar/\Msps\propto \Msps^{0.14}$
  (left panel), or equivalently $\Mstar/\Msps\propto \sigmae^{0.33}$
  (right panel), i.e., ``heavier'' IMFs in higher ``mass''
  galaxies. The tilt of the stellar mass fundamental plane enables us
  to break this degeneracy. Acceptable models are indicated by thick
  solid black lines. The best models are mass follows light at low
  masses $\Msps \lta 10^{11.2}$, and expansion or uncontracted NFW
  haloes for $\Msps \gta 10^{11.3}$.  At the highest masses halo
  contraction is allowed given reasonable uncertainties in stellar
  mass errors. Changing the stellar anisotropy from $\beta=0$
  (isotropic orbits) to $\beta=0.5$ only lowers the derived stellar
  masses by $\simeq 10\%$, as indicated by the error bar in the top
  left corner, (it does not change the conclusions regarding halo
    response). Our results are consistent with a number of recent
    studies as referenced by the caption in the lower right corner of
    each panel (see text for further details). In the left panel the
    upper horizontal axis shows the average aperture stellar velocity
    dispersions at the corresponding stellar masses.}
\label{fig:deltaIMF}
\end{figure*}

A summary of the main results of this paper, together with a
comparison to some recent results in the literature is given in
Fig.~\ref{fig:deltaIMF}. The various models with fixed halo responses
are given by colored lines (with the same color and line type as in
Figs.~\ref{fig:vm_chabrier} \& \ref{fig:dm}). The parameters of the
relations between IMF mismatch parameter and SPS mass are given in
Table ~\ref{tab:models}. The corresponding relations between IMF
mismatch parameter and stellar velocity dispersion (measured within
the effective radius) are given in Table~\ref{tab:IMFsigma}. The
models that reproduce the fundamental plane constraints (from
Fig.~\ref{fig:dvmr_model_all}) are highlighted with thick black lines.

In the left panel we compare our results with those of Auger \etal
(2010a), and Dutton \etal (2012, 2013) who directly compared true
stellar masses with SPS masses. All three of these results are in
excellent agreement with ours. The result from the study of strong
gravitational lenses by Auger \etal (2010a) is shown as a green cross
(we show their best fitting model which assumes NFW haloes).  The
vertical line indicates the 1$\sigma$ uncertainty on $\Mstar/\Msps$,
the width of the diagonal line corresponds to the observed scatter in
SPS masses, while the slope of the diagonal line shows the best
fitting slope of the relation between $\Mstar/\Msps$ and SPS mass. The
result from a study of the densest early-type galaxies in the SDSS by
Dutton \etal (2012) is shown as a magenta point (this study assumes
MFL models and justifies this based on the fact these galaxies follow
the virial fundamental plane). The lower error bar shows the effect of
changing the stellar anisotropy from $\beta=0.0$ to $\beta=0.5$. The
result for bulge IMFs from a study of massive spiral galaxy strong
lenses from Dutton \etal (2013) is given by the black square. The
error bar corresponds to the 1$\sigma$ uncertainty.

In the right panel of Fig.~\ref{fig:deltaIMF} we compare our results
with Treu \etal (2010), Conroy \& van Dokkum (2012) and Cappellari
\etal (2012b) who compared the IMF mismatch with stellar velocity
dispersion. These three results are in broad agreement with ours. The
larger differences compared with the left panel are likely due to
systematic errors in the conversion between SPS masses and velocity
dispersion definitions. We detail the corrections we applied
below. The results from the study of strong gravitational lenses by
Treu \etal (2010) are shown with green squares (this study assumes NFW
haloes with fixed scale radii). Here we have binned the individual
measurements, so that the error bars correspond to the error on the
mean. The slope inferred by Treu \etal (2010) is significantly steeper
than what we favor, although the mean is in excellent agreement. The
results from the study of stellar absorption lines by Conroy \& van
Dokkum (2012) are shown with black circles. Again here we have binned
the individual measurements, so the error bars correspond to the error
on the mean. We have converted their SPS masses from a Kroupa IMF into
a Chabrier IMF by subtracting 0.035 dex. We have converted the
aperture for the velocity dispersions from one eighth an effective
radius to one effective radius by subtracting 0.06 dex.  The results
from the study of ATLAS3D galaxies by Cappellari \etal (2012b) are
given by the thin black line. We have converted their SPS masses from
a Salpeter IMF to a Chabrier IMF by subtracting 0.23 dex. The grey
shaded region corresponds to the 1$\sigma$ uncertainty on their
fit. Their slope is in excellent agreement with ours ($\simeq 1/3$),
but there is a zero point offset of $\simeq 20\%$ compared to our best
fitting MFL model.

\begin{table}
 \centering
 \caption{Parameters of fits to IMF mismatch parameter vs velocity dispersion: $\DeltaIMF = a + b \log(\sigmae/[130\kms])$}
  \begin{tabular}{cccc}
    \hline
    \hline  
    Halo Response & Anisotropy & $\Delta_{\rm IMF}$ zero & $\Delta_{\rm IMF}$ slope \\
            $\nu$ & $\beta$    & $a$             & $b$ \\
    \hline
   1.0  & 0.0 & $-0.049\pm 0.001$ & $0.364 \pm 0.003$\\
   0.5  & 0.0 & $0.084 \pm 0.001$ & $0.353 \pm 0.002$\\
   0.0  & 0.0 & $0.164 \pm 0.001$ & $0.337 \pm 0.002$\\
 $-$0.5 & 0.0 & $0.208 \pm 0.001$ & $0.316 \pm 0.002$\\
    MFL & 0.0 & $0.215 \pm 0.001$ & $0.298 \pm 0.004$\\
    MFL & 0.5 & $0.176 \pm 0.001$ & $0.299 \pm 0.004$\\
\hline
\hline
\label{tab:IMFsigma}
\end{tabular}
\end{table}

\section{Discussion of systematic effects} 
In this section we discuss how the simplifying assumptions of our
model might lead to systematic biases in our results. We break these
down into the following areas: dynamical models; stellar masses;
and fundamental plane models.

\subsection{Dynamical Models}
We use spherical Jeans models to predict aperture velocity dispersions
of early-type galaxies.  However, early-type galaxies are in general
not spherical, with average axis ratios decreasing (i.e., becoming
flatter) with decreasing mass (e.g., Holden \etal 2012). We thus
ignore any flattening induced by anisotropy or rotation in our models.
Several studies have shown that spherical Jeans models do in
fact recover accurate dynamical masses of non-spherical systems. For
example, Cappellari \etal (2006) have shown that two and three
integral models yield dynamical masses that are consistent with those
obtained from a simple virial relation $M_{\rm dyn} \propto
\sigma_{\rm e}^2 R_{\rm e}$ for both fast- and slow-rotators. When
combining strong lensing with kinematics to determine the slope of the
total mass density profile, identical results (within the measurement
errors) are obtained from two integral dynamical models with integral
field kinematic data (Barnab{\'e} \etal 2011) and spherical Jeans
models with a single kinematic measurement (Auger \etal 2010b).  

For our fiducial models we assume isotropic velocity dispersions
($\beta=0$), which is a good approximation for massive elliptical
galaxies (e.g., Gerhard \etal 2001). Radially anisotropic velocity
dispersions ($\beta > 0$) will result in higher $\sigmaap$ and thus
lower stellar mass normalizations (see
Fig.~\ref{fig:vm_model_MFL}). In terms of the fundamental plane,
$\beta=0.5$ results in a weaker correlation by $\simeq 0.05$ than that
for isotropic orbits, and thus will not be able to reconcile models
with significant amounts of dark matter with the observed fundamental
plane tilt.

\subsection{Stellar Masses} The masses from SPS models are subject to
a number of systematic uncertainties -- from the treatment of
uncertain aspects of stellar evolution, to the star formation and
chemical enrichment histories of galaxies (e.g., Conroy \etal 2009;
Gallazzi \& Bell 2009). In principle these uncertainties could account
for the factor of $\sim 2$ deviation in stellar masses from a Milky
Way IMF that we infer. In practice, for old stellar populations, the
effects of using different existing SPS models are of order 0.05-0.1
dex (e.g., Treu \etal 2010), and thus are not enough to reconcile our
results with a universal IMF. As a specific example, a comparison of
the stellar masses that we use in this study (from the MPA/JHU
database) with those from Mendel \etal (2013) yields a systematic mass
independent offset of $\simeq 0.04$ dex (MPA/JHU masses are lower) and
a random uncertainty of $\simeq 0.08$ dex. These masses are based on
both different SPS models -- Bruzual \& Charlot (2003) vs Conroy \&
Gunn (2010), and different photometry -- SDSS vs Simard \etal (2011).
Finally, we note that even if our conclusions regarding the IMF are
biased by systematic uncertainties in SPS masses, our conclusions
regarding the true stellar masses, dark matter fractions, and dark
halo responses are independent of these uncertainties in SPS masses.

When parametrizing the deviation of the true stellar mass from the SPS
stellar mass, $\DeltaIMF=\Mstar/\Msps$, we have only considered
variations as a function of $\Msps$.  This is driven by our decision
to express scaling relations of galaxy properties, such as velocity
dispersions and sizes, as a function of $\Msps$, as well as defining
the tilt of the fundamental plane in terms of the correlation between
$\Delta \log \sigmaap | \Msps$ and $\Delta \log \Re | \Msps$. In this
context, if we were to consider $\DeltaIMF$ variation as a function of
velocity dispersion, as has been suggested by numerous authors (e.g.,
Treu \etal 2010; Conroy \& van Dokkum 2012; Cappellari \etal 2012b),
we should recalculate the observed tilt of the fundamental plane for
each variation considered, which would greatly increase the complexity
of the model. However, since velocity dispersion and stellar mass are
tightly correlated (e.g., see Fig.~\ref{fig:vmr}), these effects are
small, and to first order any variation of $\DeltaIMF$ with velocity
dispersion will also result in a equivalent variation with $\Msps$.
Fig.~\ref{fig:deltaIMF} indeed shows that in our models the IMF
mismatch parameter is correlated with velocity dispersion, with a
slope of $\simeq 0.33$ (see Table~\ref{tab:IMFsigma} for parameters of
the fits). Our results are also consistent with results from previous
measurements (e.g., Treu \etal 2010; Conroy \& van Dokkum 2012;
Cappellari \etal 2012b). As discussed above, the small (of order
10-20\%) differences between the various measurements are consistent
with systematic uncertainties in SPS masses and conversions between
masses using different IMFs.

\subsection{Fundamental Plane Models}
When making model predictions for the tilt of the fundamental plane we
assume (for simplicity) that variation in galaxy sizes at fixed
stellar mass are independent of the properties of the dark matter halo
(e.g., concentration, halo response parameter, halo mass). As we have
shown above, with this assumption, models with halo contraction (and
Milky Way type IMFs) predict a weaker fundamental plane tilt than
observed. In order for such models to reproduce the observed
fundamental plane tilt would require smaller galaxies (at fixed
stellar mass) to have higher dark matter masses within an effective
radius, and vice versa for larger galaxies. (We note that an
interesting consequence of such a correlation is to reduce the scatter
in the dark matter fractions within an effective
radius). Qualitatively this could be achieved by introducing an
anti-correlation between the size of a galaxy and the concentration of
the dark matter halo, halo response parameter $\nu$, or halo mass.
Determining whether any of these correlations are expected in a \LCDM
context could be possible using cosmological hydrodynamical
simulations, or semi-analytic models.

Strong gravitational lensing enables the slope of the total mass
density profile to be measured within roughly the effective radius
(e.g., Koopmans \etal 2006). This slope is generally referred to as
$\gamma'$. Auger \etal (2010a) used this information implicitly to
constrain the IMF and halo response. In Dutton \& Treu (2013) the
observed distribution of $\gamma'$ is compared to that of a sub-sample
of our models (with velocity dispersions $\sigmaap \sim 250\pm 40
\kms$). Remarkably, the uncontracted NFW model, which best matches the
fundamental plane tilt data at high masses (upper right panel of
Fig.~\ref{fig:dvmr_model_all}), also best matches the observed
distribution of $\gamma'$. Not only does this provide independent
confirmation of our model assumptions: i.e., that variation in galaxy
sizes is uncorrelated with variation in halo masses and
concentrations, it also shows that the observed non-homology in total
mass density profiles can be explained by a relatively simple model.


\section{Summary}
\label{sec:sum}

We use the relations between aperture stellar velocity dispersion
($\sigmaap$), stellar mass ($\Msps$), and galaxy size ($\Re$) to
constrain the dark halo response and stellar IMF in early-type
galaxies. Our observational sample consists of $\sim$ 150 000
early-type galaxies from the SDSS/DR7 at redshift $z\sim 0.1$.  We
construct mass models with \LCDM haloes (modified by contraction or
expansion), as well as mass-follows-light (MFL), that reproduce the
observed distribution of SPS stellar masses and galaxy sizes.  Our
models include uncorrelated scatter in galaxy sizes, halo masses and
halo concentrations.  The remaining free parameters of the model are
constrained by the median relation between $\sigmaap$ and $\Msps$ (the
Faber-Jackson relation), and the strength of the correlation between
residuals of the $\sigmaap$ vs $\Msps$ and effective radius $\Re$ vs
$\Msps$ relations: $\dvr \equiv \Delta\log\sigmaap/\Delta\log\Re$
(equivalent to the tilt of the fundamental plane).

Our constraints on the stellar IMF are obtained by comparing the true
stellar masses (which we derive with our dynamical models) with those
derived from stellar population synthesis models. Thus all of our
conclusions regarding the IMF are dependent on there being no large
systematic errors in SPS masses. Our constraints on the halo response
are however independent of these uncertainties.
We summarize our results in Fig.~\ref{fig:deltaIMF} and as follows:

\begin{itemize}

\item Reproducing the median $\sigmaap-\Msps$ relation is not possible
  for models with {\it both} a universal IMF and universal dark halo
  response.  Significant departures in either/or both are required.

\item Models with a universal halo response to galaxy formation (as
  well as MFL) require heavier IMFs in higher mass galaxies and are
  consistent with $\Mstar/\Msps \propto \Msps^{0.14} \propto
  \sigmae^{0.33}$.  At a stellar mass of $\Msps=10^{11}\Msun$ these
  models with adiabatic contraction require close to Chabrier IMFs,
  models with uncontracted NFW haloes require close to Salpeter IMFs,
  while MFL models require IMFs heavier than Salpeter.

\item A model with a universal (close to Kroupa) IMF and mass
  dependent dark halo response ranging from expansion at low masses
  ($\Msps \lta 10^{10}\Msun$) to adiabatic contraction at high masses
  ($\Msps \gta 10^{11}\Msun$) is able to reproduce the
  $\sigmaap-\Msps$ relation.

\item We find that $\dvr$ varies systematically with stellar mass. The
  minimum is $\dvr \simeq -0.45$ at $\Msps=10^{10}\Msun$ (Chabrier
  IMF) and increases to $\dvr \simeq -0.15$ at $\Msps=10^{11.6}\Msun$.
  The virial fundamental plane has $\dvr= -1/2$. And thus the observed
  tilt of the fundamental plane is not a constant, as is generally
  assumed.

\item The MFL model successfully reproduces the mass dependent
  fundamental plane tilt for masses $\Msps \lta 10^{11.2}\Msun$.
  However, above $\Msps \sim 10^{11.2}$ the MFL model progressively
  over-predicts the strength of $\dvr$. Reproducing the observations
  is possible with halo expansion for masses $\Msps \gta 10^{11.2}$,
  and uncontracted NFW haloes for $\Msps \gta 10^{11.3}$. At the
  highest masses models with halo contraction are also consistent with
  the data.

\item Models with a universal IMF (and non-universal halo response)
  are unable to reproduce the tilt of the fundamental plane.

\end{itemize}

Our results are in agreement with several recent studies which also
favor IMFs that are ``heavier'' than Salpeter in massive galactic
spheroids, in addition to an effective stellar mass or velocity
dispersion dependence to the IMF (van Dokkum \& Conroy 2010; Auger
\etal 2010a; Sonnenfeld \etal 2012; Conroy \& van Dokkum 2012;
Cappellari \etal 2012b; Newman \etal 2013; Dutton \etal 2013).  As a
final remark, we note that the absence of adiabatic contraction
implied by our models indicates that non-dissipative mergers and/or
feedback play an important role in the formation of early-type
galaxies of {\it all} masses.

\section*{Acknowledgments} 

We thank the anonymous referee for a constructive report that helped
to improve the clarity of the manuscript. We thank Eric Bell, Charlie
Conroy, and Michele Cappellari for valuable discussions.

AVM acknowledges financial support to the DAGAL network from the
People Programme (Marie Curie Actions) of the European Union's Seventh
Framework Programme FP7/2007-2013/ under REA grant agreement number
PITN-GA-2011-289313.

Funding for the Sloan Digital Sky Survey (SDSS) has been provided by
the Alfred P. Sloan Foundation, the Participating Institutions, the
National Aeronautics and Space Administration, the National Science
Foundation, the U.S.  Department of Energy, the Japanese
Monbukagakusho, and the Max Planck Society.  The SDSS Web site is
http://www.sdss.org/.

The SDSS is managed by the Astrophysical Research Consortium (ARC) for
the Participating Institutions. The Participating Institutions are The
University of Chicago, Fermilab, the Institute for Advanced Study, the
Japan Participation  Group, The  Johns Hopkins University,  Los Alamos
National  Laboratory, the  Max-Planck-Institute for  Astronomy (MPIA),
the  Max-Planck-Institute  for Astrophysics  (MPA),  New Mexico  State
University, University of Pittsburgh, Princeton University, the United
States Naval Observatory, and the University of Washington.


\appendix

\section{constrained $\Lambda$CDM based mass models of early-type
  galaxies}
\label{sec:mm}
In this appendix we describe the mass models we construct to reproduce
the observed structural and dynamical scaling relations of early-type
galaxies. These are the same as described in Dutton \etal (2011a).
Our most general mass model consists of three spherical components: a
stellar component with S\'ersic index $n=4$; a stellar component with
S\'ersic index $n=1$; and a dark matter halo. The dark matter halo is
an NFW (Navarro, Frenk, \& White 1997) modified by the response of the
halo to galaxy formation.  This model has 7 parameters (4 for the
stars and 3 for the dark matter), 5 of which are determined by
observations and theory. The distribution of stellar mass in galaxies
is described by three relations: half-light size vs SPS mass (for the
$n=1$ and $n=4$ components) and de Vaucouleurs fraction $f_{\rm deV}$
vs SPS mass (the fits to these relations are given in
Table~\ref{tab:fits}).  The relation between SPS mass and dark halo
mass is taken from the compilation of observations by Dutton \etal
(2010). The relation between dark halo concentration and dark halo
mass is taken from \LCDM N-body simulations in a WMAP5 cosmology
(Macci\`o \etal 2008).

The two unknowns are the normalization of the stellar masses, which we
term $\Delta_{\rm IMF}$, and the response of the dark matter halo to
galaxy formation.  Following Dutton \etal (2007; 2011) we consider a
range of halo responses parametrized by the parameter $\nu$. 
Where standard adiabatic contraction (Blumenthal \etal 1986)
corresponds to $\nu=1$, the contraction model of Gnedin \etal (2004)
can be approximated with $\nu\simeq 0.8$, the contraction model of
Abadi \etal (2010) can be approximated with $\nu\simeq 0.4$, an
unmodified halo corresponds to $\nu=0$, while expansion corresponds to
$\nu < 0$. As a limiting case of maximum halo expansion we also
consider models in which mass follows light (i.e., no dark matter).

The combinations of allowed $\DeltaIMF$ and $\nu$ are constrained by
the observed velocity dispersion vs SPS mass relation
(Fig.~\ref{fig:vmr}, Table ~\ref{tab:datavm}). In Dutton \etal (2011a)
we used empirical constraints (from strong lensing and dynamics) for
the average conversion between circular velocity at the half-light
radius $V_{\rm circ}(\Re)$ and the velocity dispersion within the
half-light radius, $\sigma_{\rm e}$. In this paper we compute aperture
velocity dispersions for our models by solving the spherical Jeans
equations, as describe below.

\begin{table*}
 \centering
 \caption{Observed aperture velocity dispersion ($\sigma_{\rm ap}$) - stellar mass ($M_{\rm SPS}$) and circularized effective radius ($R_{\rm e}$) - stellar mass relations from Fig.~\ref{fig:vmr}, and the correlation between the residuals of these relations from Fig.~\ref{fig:dvr_mass}}
  \begin{tabular}{cccccccccc}
    \hline
    \hline  
\multicolumn{10}{l}{$\log_{10}(M_{\rm SPS}/\Msun)$ 
\hspace{1.0cm} $\log_{10}(\sigma_{\rm ap}/\kms)$ 
\hspace{3.0cm} $\log_{10}( R_{\rm e}/\kpc)$
\hspace{3.0cm} $\Delta\log_{10}(\sigma_{\rm ap})/\Delta\log_{10}(R_{\rm e})$}\\
min & max &  median & median & 16th \%\,ile & 84th \%\,ile & median  & 16th \%\,ile & 84th \%\,ile\\
    \hline
 9.4 & 9.6 &  9.515 &  1.748$\pm$0.009 & 1.602 & 1.856 & 0.078$\pm$0.012 & $-$0.133 & 0.236 & $-$0.368$\pm$0.027($^{+0.134}_{-0.138}$)\\ 
 9.6 & 9.8 &  9.712 &  1.822$\pm$0.005 & 1.710 & 1.942 & 0.083$\pm$0.009 & $-$0.106 & 0.256 & $-$0.385$\pm$0.018($^{+0.096}_{-0.064}$)\\ 
 9.8 &10.0 &  9.912 &  1.896$\pm$0.004 & 1.786 & 2.028 & 0.139$\pm$0.006 & $-$0.051 & 0.313 & $-$0.443$\pm$0.012($^{+0.051}_{-0.024}$)\\ 
10.0 &10.2 & 10.112 &  1.977$\pm$0.002 & 1.862 & 2.091 & 0.199$\pm$0.004 &  0.012 & 0.361 & $-$0.406$\pm$0.009($^{+0.015}_{-0.017}$)\\ 
10.2 &10.4 & 10.313 &  2.056$\pm$0.001 & 1.955 & 2.153 & 0.260$\pm$0.002 &  0.106 & 0.425 & $-$0.377$\pm$0.006($^{+0.011}_{-0.007}$)\\ 
10.4 &10.6 & 10.508 &  2.119$\pm$0.001 & 2.031 & 2.208 & 0.348$\pm$0.002 &  0.206 & 0.498 & $-$0.357$\pm$0.004($^{+0.009}_{-0.005}$)\\ 
10.6 &10.8 & 10.707 &  2.179$\pm$0.001 & 2.093 & 2.261 & 0.453$\pm$0.001 &  0.324 & 0.592 & $-$0.337$\pm$0.004($^{+0.004}_{-0.003}$)\\ 
10.8 &11.0 & 10.903 &  2.238$\pm$0.001 & 2.155 & 2.310 & 0.564$\pm$0.001 &  0.450 & 0.691 & $-$0.304$\pm$0.003($^{+0.003}_{-0.002}$)\\ 
11.0 &11.2 & 11.098 &  2.294$\pm$0.001 & 2.218 & 2.359 & 0.676$\pm$0.001 &  0.568 & 0.796 & $-$0.261$\pm$0.003($^{+0.002}_{-0.001}$)\\ 
11.2 &11.4 & 11.293 &  2.347$\pm$0.001 & 2.280 & 2.405 & 0.791$\pm$0.001 &  0.688 & 0.908 & $-$0.197$\pm$0.003($^{+0.005}_{-0.004}$)\\ 
11.4 &11.6 & 11.486 &  2.392$\pm$0.001 & 2.333 & 2.447 & 0.921$\pm$0.001 &  0.814 & 1.042 & $-$0.152$\pm$0.004($^{+0.006}_{-0.004}$)\\ 
11.6 &11.8 & 11.678 &  2.431$\pm$0.001 & 2.375 & 2.482 & 1.064$\pm$0.002 &  0.951 & 1.181 & $-$0.148$\pm$0.007($^{+0.000}_{-0.002}$)\\ 
\hline
\hline
\label{tab:datavm}
\end{tabular}
\end{table*}

\subsection{Predicting SDSS aperture velocity dispersions.}
\label{sec:nonhomology}
As described in the previous section, we have spherical models for the
distribution of total and stellar mass. Given an assumption of the
anisotropy profile we can then solve the spherical Jeans equations to
get the radial velocity dispersion profile. We can then compute the
projected velocity dispersions inside the SDSS aperture including
  the effects of seeing. The relevant equations are given below.

We consider spherical galaxy models with 3D stellar density
distribution $\rho_*(r)$, and projected stellar density distribution
$\Sigma_*(R)$. The radial component $\sigma_r(r)$ of the velocity
dispersion tensor is found by solving the Jeans equations:
\begin{equation}
  \frac{ d (\rho_* \sigma_r^2)}{dr} + \frac{2\beta}{r} \rho_*\sigma_r^2 = -\rho_* \frac{G M(r)}{r^2},
\end{equation}
where $\beta$ is the velocity anisotropy, and $M(r)$ is the
spherically enclosed mass within radius $r$.
The solution of the Jeans equation is given by
\begin{equation}
\rho_*\sigma_r^2 = \frac{1}{I}\int_r^{\infty} I \rho_* \frac{G M}{x^2} dx
\end{equation}
where $I= \exp \int (2\beta/r)dr$ is the integrating factor. For
constant $\beta$ the integrating factor is $I = r^{2\beta}$
If we start with a 2D surface density profile (e.g., a
S\'ersic profile), the 3D surface density profile is given by
\begin{equation}
\label{eq:sigma}
  \rho_*(r)= -\frac{1}{\pi}\int_r^{\infty} \frac{d\Sigma_*(R)}{dR} \frac{ dR}{\sqrt{R^2-r^2}}.
\end{equation}
Alternatively, if we start with a 3D density profile (e.g., a
Hernquist profile which approximates a S\'ersic $n=4$ profile), the
projected density profile is
\begin{equation}
\label{eq:rho}
  \Sigma_*(R)= 2 \int_R^{\infty} \rho_*(r) \frac{ r}{\sqrt{r^2-R^2}} \,dr.
\end{equation}
In practice, for our default models, we adopt two components: an
exponential (in projection) for which Eq.~\ref{eq:sigma} has an
analytic solution (e.g., van den Bosch \& de Zeeuw 1996), and a
Hernquist profile for which Eq.~\ref{eq:rho} has an analytic solution
(Hernquist 1990). The parameters of these two components that we use
in our models are given in Table~\ref{tab:fits}.

The projected velocity dispersion is given by
\begin{equation}
  \Sigma_* \sigma_p^2 = 2 \int_R^{\infty}\left[1-\beta\frac{R^2}{r^2} \right]\rho_*\sigma_r^2 \frac{r}{\sqrt{r^2-R^2}}\,dr.
\end{equation}
The SDSS spectra are measured within a 3 arcsec diameter aperture, so the aperture projected velocity dispersion is given by
\begin{equation}
\sigma_{\rm ap}^2 = \frac{ \int^{R_{\rm ap}}_0 \langle \Sigma_*\,\sigma_p^2 \rangle \, 2\pi \,R \,dR }{ \int_0^{R_{\rm ap}}\langle \Sigma_* \rangle \,2\pi \,R \,dR }.
\end{equation} 
Where the angled brackets indicate variables that have been convolved
with the seeing. We assuming a Gaussian with FWHM=1.4 arcsec, which is
the median seeing for SDSS (Simard \etal 2011).

\begin{table*}
 \centering
 \caption{Parameters of the fitting formula (Eqs.~\ref{eq:power2} \& 
\ref{eq:s2}) for various scaling relations used in this paper.}
 \begin{tabular}{cccccccc}
\hline
\hline  
$y$ & $x$ &  $\alpha$ & $\beta$ & $x_0$ & $y_{0}$ & $\gamma$ \\
\hline
$\log_{10}(\sigma_{\rm ap}/\kms)$ & $\log_{10}(\Msps/\Msun)$ & 0.42 & 0.20 & 10.7 & 2.18 & 1.0  \\
$\log_{10}(R_{\rm e}/\kpc)$       & $\log_{10}(\Msps/\Msun)$ & 0.00 & 0.70 & 10.2 & 0.22 & 1.0 \\
$\log_{10}(R_{\rm ap}/\kpc)$      & $\log_{10}(\Msps/\Msun)$ & 0.45 & 0.24 & 11.1 & 0.54 & 1.2  \\
$\sigma_{\log R_{\rm ap}}$          & $\log_{10}(\Msps/\Msun)$ & 0.046 & -0.037 & 10.1 & 0.098 & 2.0 \\
$f_{\rm deV}$                     & $\log_{10}(\Msps/\Msun)$ & 0.17 & -0.22 & 10.9 & 0.65 & 2.0   \\
$\log_{10}(R_{\rm deV}/\kpc)$     & $\log_{10}(\Msps/\Msun)$ & -0.01 & 0.55 & 10.0 & -0.08 & 2.0   \\
$\log_{10}(R_{\rm exp}/\kpc)$     & $\log_{10}(\Msps/\Msun)$ &  0.29 & 0.61 & 10.1 &  0.39 & 1.0 \\
$e \log \sigma_{\rm ap}$         & $\log_{10}(\sigma_{\rm ap}/\kms)$ & -0.8 &  0.0 & 1.7 & 0.093 & 3.8 \\
$e \log R_{\rm e}$               & $\log_{10}(R_{\rm e}/\kpc)$ & ... & ... & ... & 0.035 & ... \\
$e \log \Msps$                  & $\log_{10}(\Msps/\Msun)$ & ... & ... & ... & 0.092 & ... \\
\hline
$y$ & $x$ &  $y_1$ & $y_2$ & $x_0$ & & $\gamma$\\
$\sigma_{\log_{10}(\sigmaap)}$  & $\log_{10}(\Msps/\Msun)$ & 0.12 & 0.045 & 10.7 & ... & 1.0 \\
$\sigma_{\log_{10}(\Re)}$      & $\log_{10}(\Msps/\Msun)$ & 0.18 & 0.10 & 10.55 & ... & 1.8 \\
\hline
\label{tab:fits}
\end{tabular}
\end{table*}

\begin{figure}
\centerline{
\psfig{figure=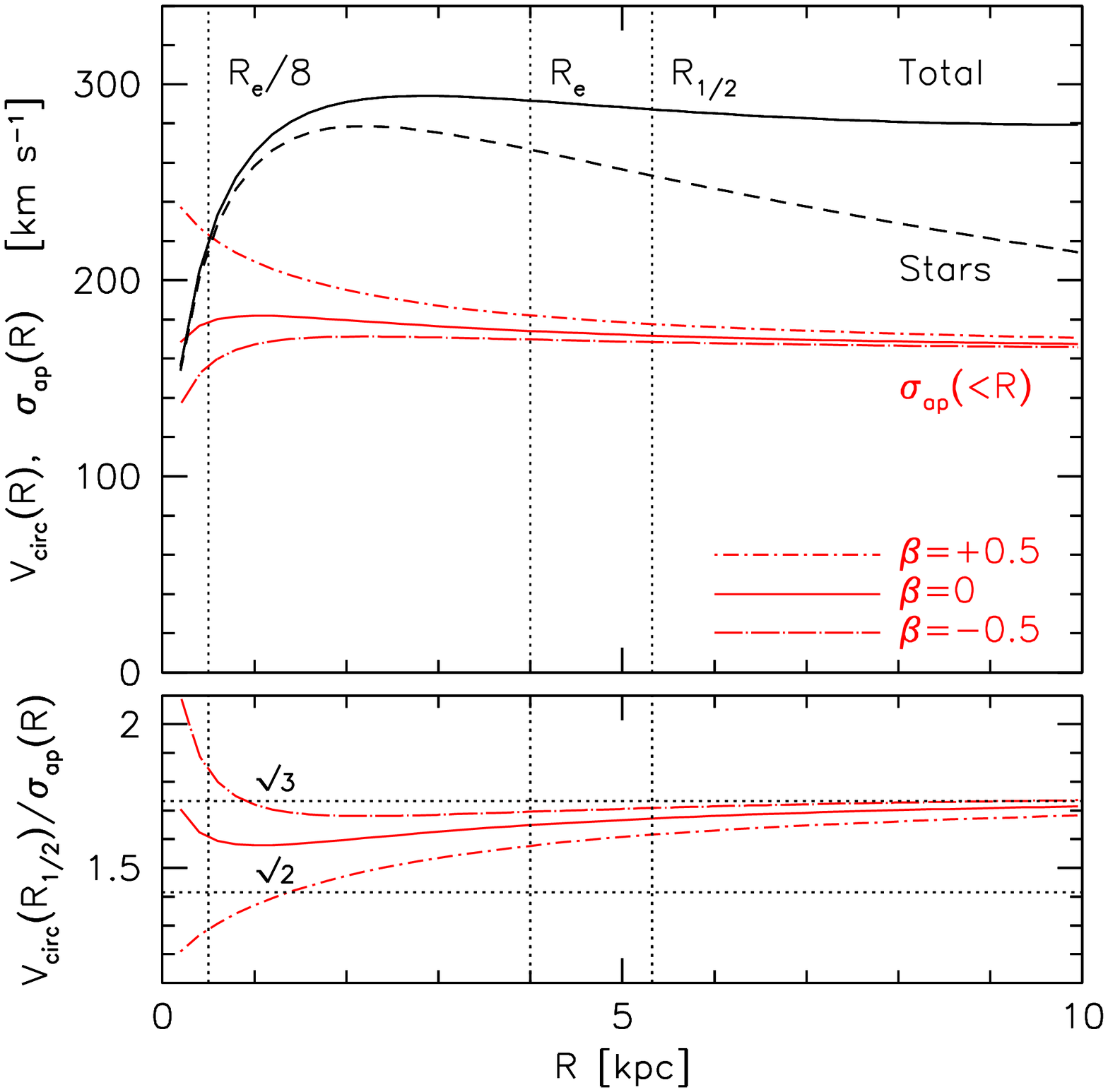,width=0.47\textwidth}
}
\caption{Example circular velocity and projected velocity-dispersion
  profiles for a model galaxy made of a Hernquist sphere of stars and
  a cosmologically motivated NFW dark matter halo.}
\label{fig:jeans}
\end{figure}

Fig.~\ref{fig:jeans} shows an example mass model consisting of an NFW
dark matter halo and a Hernquist bulge of stars. The upper panel shows
circular velocity (black lines) and aperture velocity dispersion (red
lines) profiles. The total circular velocity is given by the solid
black line, while the contribution from the stars is given by the
black dashed line. The dark matter halo makes up the difference
between these two lines.  The aperture velocity dispersions are given
for different values of the anisotropy parameter, $\beta$, chosen to
bracket the values observed in early-type galaxies (Gerhard \etal
2001).

The vertical lines mark useful apertures: $\Re$ is the projected
half-light radius (also known as the effective radius), $R_{1/2}$ is
the 3D half-light radius.  The two most common apertures for measuring
or ``correcting'' aperture velocity dispersions are $\Re$ and $\Re/8$.

As shown by (e.g., Wolf \etal 2010), if the radial velocity dispersion
profile is constant (i.e., $d\ln\sigma^2_r/d\ln r = 0$), then the mass
within the 3D half-light radius $M_{1/2}=3\sigma^2_{\rm los}
R_{1/2}/G$, where $\sigma^2{\rm los}$ is the integrated line-of-sight
velocity dispersion of the system. This can be more compactly
expressed in terms of circular velocity:
$V_{1/2}=\sqrt{3}\sigma^2_{\rm los}$. However, in general galaxies do
not have constant velocity dispersion profiles, and for galaxies in
SDSS the observed aperture typically contains less half the light (see
below).  Thus we expect some variation from the formula from Wolf
\etal (2010).

The lower panel in Fig.~\ref{fig:jeans} shows the relation between the
circular velocity at the 3D half-light radius ($V_{1/2}$), and the
projected velocity dispersion ($\sigma_{\rm ap}(R)$) measured within a
radius $R$. For large apertures $R\sim 2\Re$ the circular velocity is
indeed almost independent of anisotropy. However, for the typical
apertures for SDSS galaxies ($0.5\lta \Re \lta 1.0$), there is a
non-negligible dependence of circular velocity on
anisotropy. Furthermore, when using velocity dispersions within on
eighth of an effective radius, there is a large dependence of circular
velocity on anisotropy $V_{1/2}/\sigma_{\rm ap}$ varies from 1.3 to
1.8. Thus by correcting the aperture to smaller radii than observed
information is lost, while correcting the aperture to larger radii
than observed a specific mass profile is assumed.  This motivates us
to model the aperture velocity dispersions directly.

\begin{figure}
\centerline{
\psfig{figure=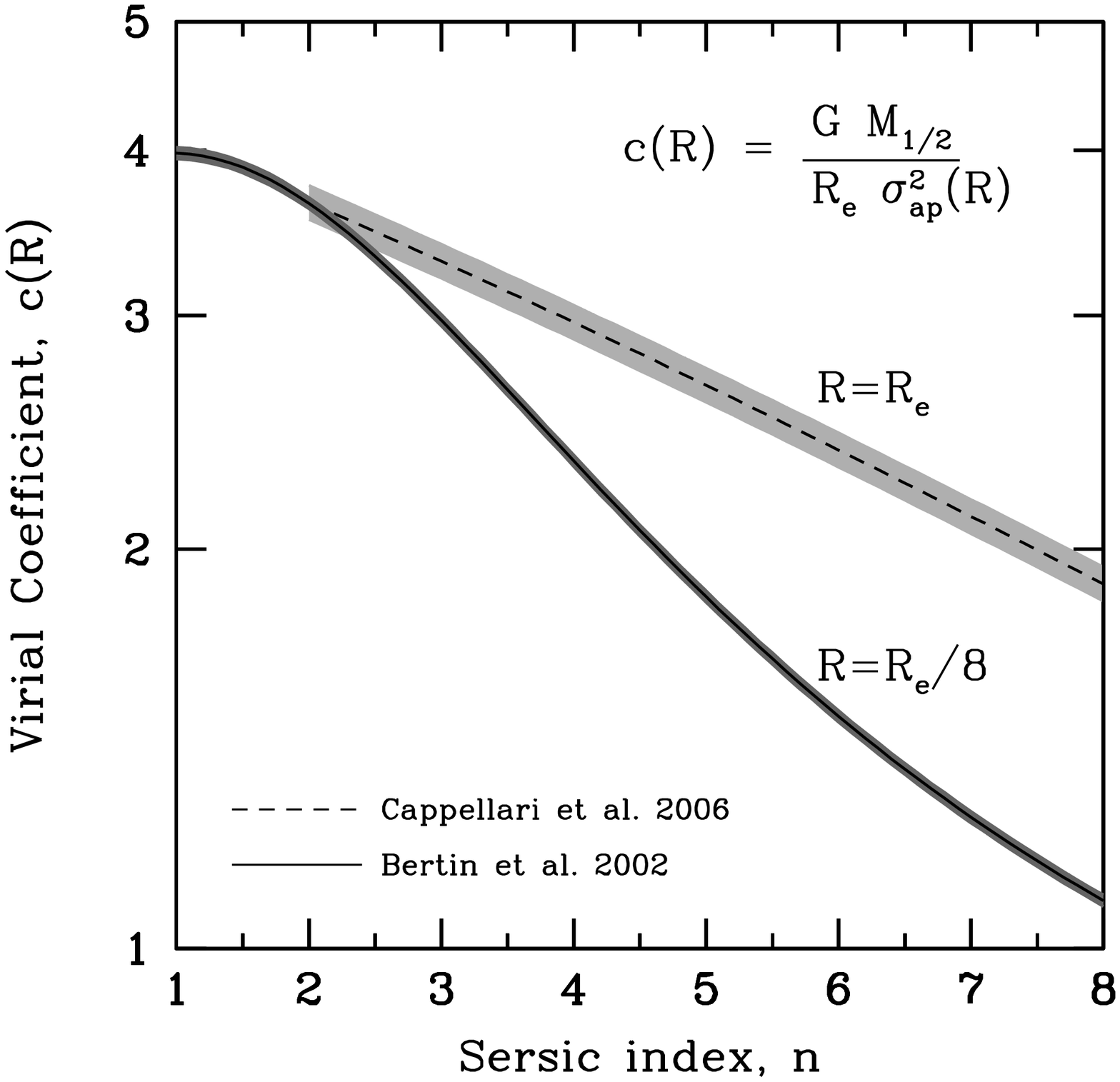,width=0.47\textwidth}
}
\caption{Virial coefficient (assuming isotropy and mass-follows-light)
  as a function of S\'ersic index for velocity dispersions measured
  within the effective radius, $R_{\rm e}$ (using the relation from
  Cappellari \etal 2006), and one eighth of the effective radius (using
  the relation from Bertin \etal 2002). The shaded regions show the
  maximum quoted error in the fitting formulae. The effect of
  non-homology on dynamical masses is weaker when velocity dispersions
  are measured within larger apertures.}
\label{fig:nonhomology}
\end{figure}

\begin{figure}
\centerline{
\psfig{figure=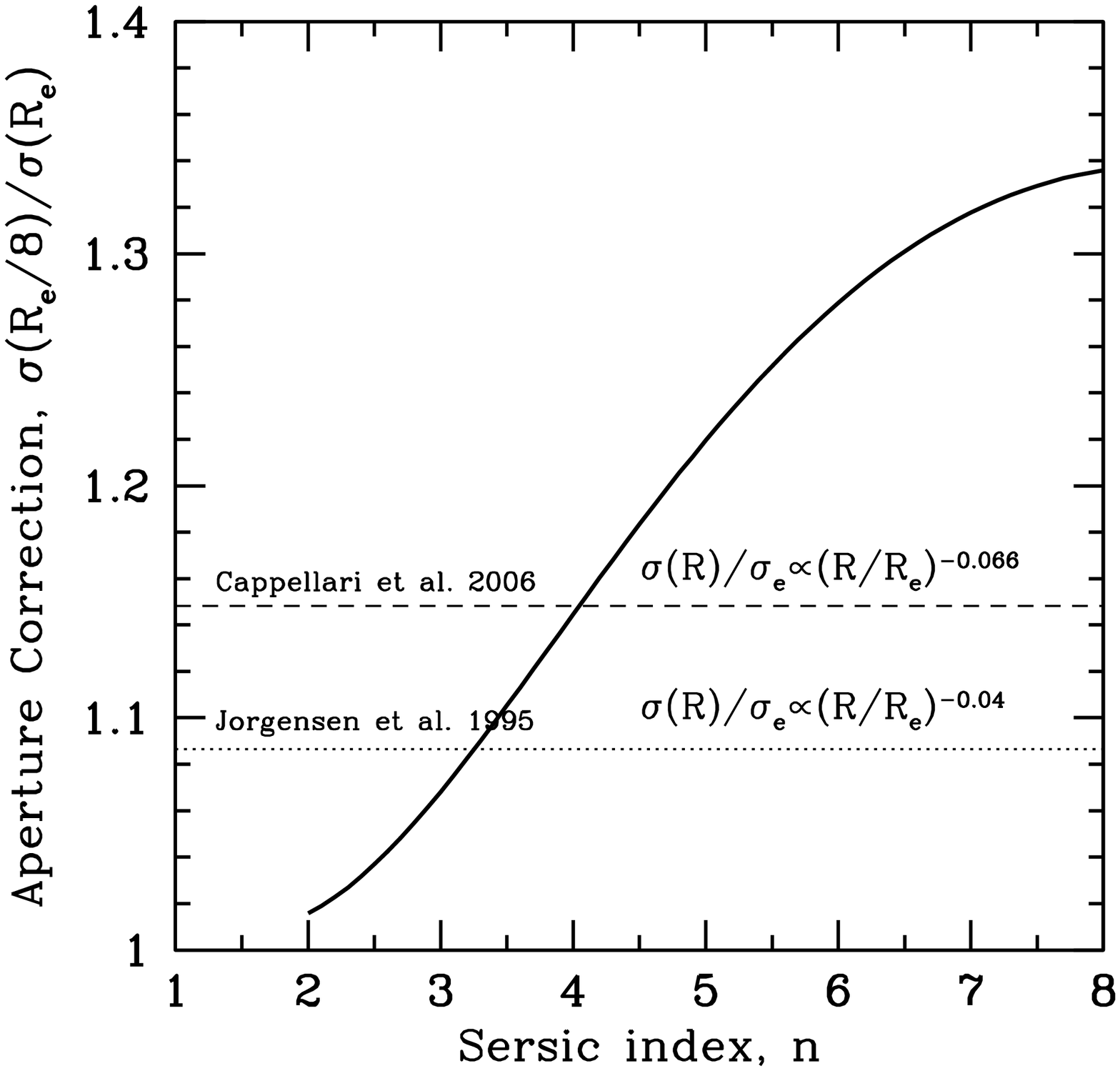,width=0.47\textwidth}
}
\caption{Aperture correction as a function of S\'ersic index as
  implied by mass follows light models with isotropic stellar orbits
  from Fig.~\ref{fig:nonhomology}. The standard aperture corrections
  (Jorgensen \etal 1995 - dotted line; Cappellari \etal 2006 -- dashed
  line) are independent of S\`ersic index, and thus inconsistent with
  mass-follows-light models with high ($n \gta 4$) or low ($n\lta 3$)
  S\'ersic indices. }
\label{fig:aperture}
\end{figure}

The aperture one measures the velocity dispersions inside also impacts
the strength of non-homology on the derived dynamical masses.  Using
relations from Bertin \etal (2002) and Cappellari \etal (2006)
Fig.~\ref{fig:nonhomology} shows that effects of non-homology on
dynamical masses are significantly weaker when velocity dispersions
are measured within the effective radius (as is typical of galaxies in
our study) compared to one eighth of the effective radius (as is often
used in the literature e.g., Trujillo \etal 2004; Taylor \etal 2010).

Since the standard aperture correction to fiber velocity dispersions
(e.g., Jorgensen \etal 1995) results in a constant offset between
velocity dispersions measured with $\Re$ and $\Re/8$,
Fig.~\ref{fig:nonhomology} reveals a fundamental
inconsistency. Obviously, the dynamical masses should be independent
of the aperture used to measure the velocity dispersion. In order for
this to be the case with mass-follows-light S\'ersic models, requires
the aperture correction depends on S\'ersic index (see
Fig.~\ref{fig:aperture}):

\begin{equation}
\sigmaei/\sigmae = \sqrt{c(\Re)/c(\Re/8)},
\end{equation}
where $c(\Re)$ is (Cappellari \etal (2006)
\begin{equation}
c(\Re) = \frac{1}{2}[8.87 -0.831n +0.0241n^2],
\end{equation}
and $c(\Re/8)$ is (Bertin \etal 2002)
\begin{equation}
c(\Re/8) = \frac{1}{2}\{73.32/[ 10.465 +(n-0.95)^2] +0.954\}.
\end{equation}
The factor of half in front of the virial coefficients is due to our
defining the dynamical mass to be with the spherical half-light
radius, rather than the total mass.

\subsection{Physical aperture sizes}

\begin{figure}
  \centerline{
    \psfig{figure=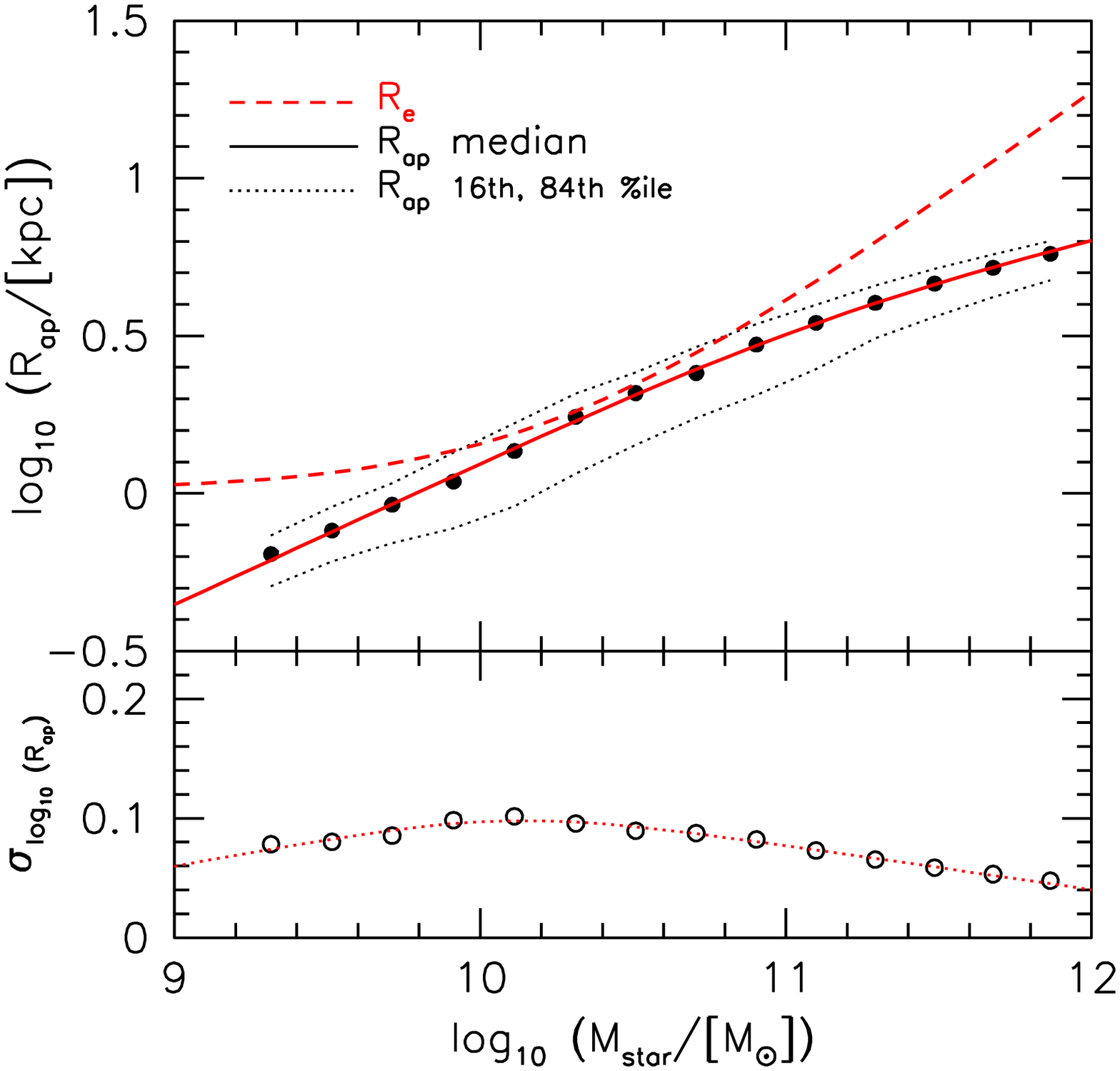,width=0.47\textwidth}
  }
\caption{Aperture size - stellar mass relation. {\it Upper panel:}
  Solid points show median of aperture sizes in kpc, solid line shows
  a fit to these points. The dotted lines show the 16th and 84th
  percentiles. For comparison the red dashed line shows the median
  half-light radius. This shows that for all but the highest and
  lowest masses, the SDSS fiber aperture, on average, corresponds to
  the half-light radius. {\it Lower panel:} Scatter in the aperture
  sizes.}
\label{fig:rap}
\end{figure}

Since we are modeling the aperture velocity dispersion, we need to
know the physical size of the aperture. This will depend on the
redshift of the galaxy. Fig.~\ref{fig:rap} shows the aperture radius
in kpc as a function of stellar mass. In the upper panel the solid
points show the median relation, while the solid line is a fit to
these data points (see Table~\ref{tab:fits} for the parameters of the
fit).  The red dashed line shows the median effective radius (from
Fig.~\ref{fig:vmr}), which shows that for all but the lowest and
highest mass galaxies the average SDSS aperture is close to the
half-light radius. For the most massive galaxies the median aperture
is roughly $0.5 \Re$. There is scatter in the aperture radius at fixed
mass, but it turns out not to induce any significant scatter to the
model velocity dispersions.

\subsection{Sampling of Galaxies}

\begin{figure}
  \centerline{ \psfig{figure=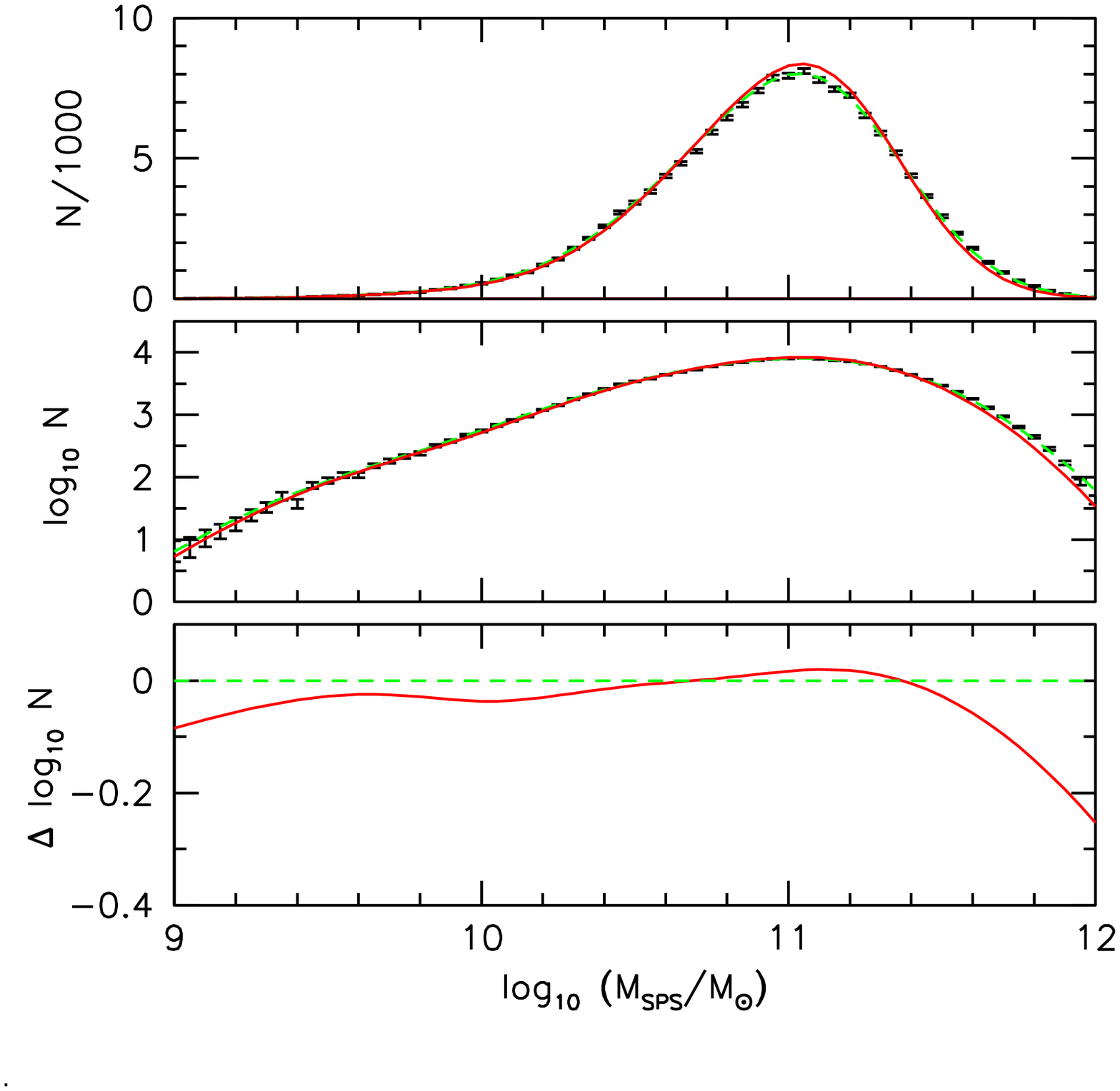,width=0.47\textwidth} }
  \caption{Observed distribution of galaxy stellar masses (assuming a
    Chabrier IMF). {\it Upper panel:} black points with error bars
    show the data, red lines show a triple Gaussian fit, green lines
    show the intrinsic distribution after correcting for measurement
    errors of 0.1 dex in stellar mass. {\it Middle panel:} Same as
    upper panel but for a logarithmic scale. {\it Lower panel:}
    Difference between $\log_{10}N$ of the model and deconvolved
    model. At high stellar masses the differences can be up to a
    factor of $\sim 1.6$. }
\label{fig:mstar}
\end{figure}

The observed sampling of the stellar masses of galaxies is not
uniform. The distribution is very roughly log-normal with a peak at
$\Msps=10^{11}\Msun$ (error bars in Fig.~\ref{fig:mstar}). Measurement
uncertainties in stellar masses will preferentially scatter galaxies
away from the peak. We fit the distribution of stellar masses with a
set of Gaussians (red lines in Fig.~\ref{fig:mstar}), and since the
measurement errors are roughly constant, we can thus deconvolve the
observed distribution of stellar masses analytically (green lines in
Fig.~\ref{fig:mstar}). The differences between the model and
deconvolved model are shown in the lower panel of
Fig.~\ref{fig:mstar}. The most significant differences occur for high
masses ($\Msps \gta 10^{11.5}\Msun$), where the deconvolved model has
up to a factor of 1.6 fewer galaxies.  The net effect of the
non-uniform stellar mass distribution coupled to errors on stellar
masses is to make the observed relations (slightly) shallower than the
true relations.

\subsection{Adding scatter to the models}

We add both intrinsic and observational sources of scatter to our
models.  There are three primary sources of intrinsic scatter: 1)
Scatter in galaxy size at fixed stellar mass; 2) Scatter in the
``pristine'' dark halo concentration at fixed halo mass; and 3)
Scatter in dark halo mass at fixed stellar mass.  The scatter in size
at fixed stellar mass is taken from the observed RM relation
(Fig.~\ref{fig:vmr}).  The scatter in the pristine halo concentration
for relaxed haloes in cosmological simulations is $\simeq 0.11$ dex
(Jing 2000; Wechsler \etal 2002; Macci\`o \etal 2008).  The scatter in
stellar mass at fixed halo mass is about 0.15 dex (e.g., More \etal
2011). However due to shallow slope of the stellar mass vs halo mass
relation at high halo masses, the scatter in halo mass at fixed
stellar mass increases with stellar mass.  The results from More \etal
(2011) for red galaxies can be approximated by $\sigma_{\log \Mvir} =
0.15 + 0.2 (\log \Msps -10.4)$, with a minimum of 0.15.

\begin{figure}
\centerline{
\psfig{figure=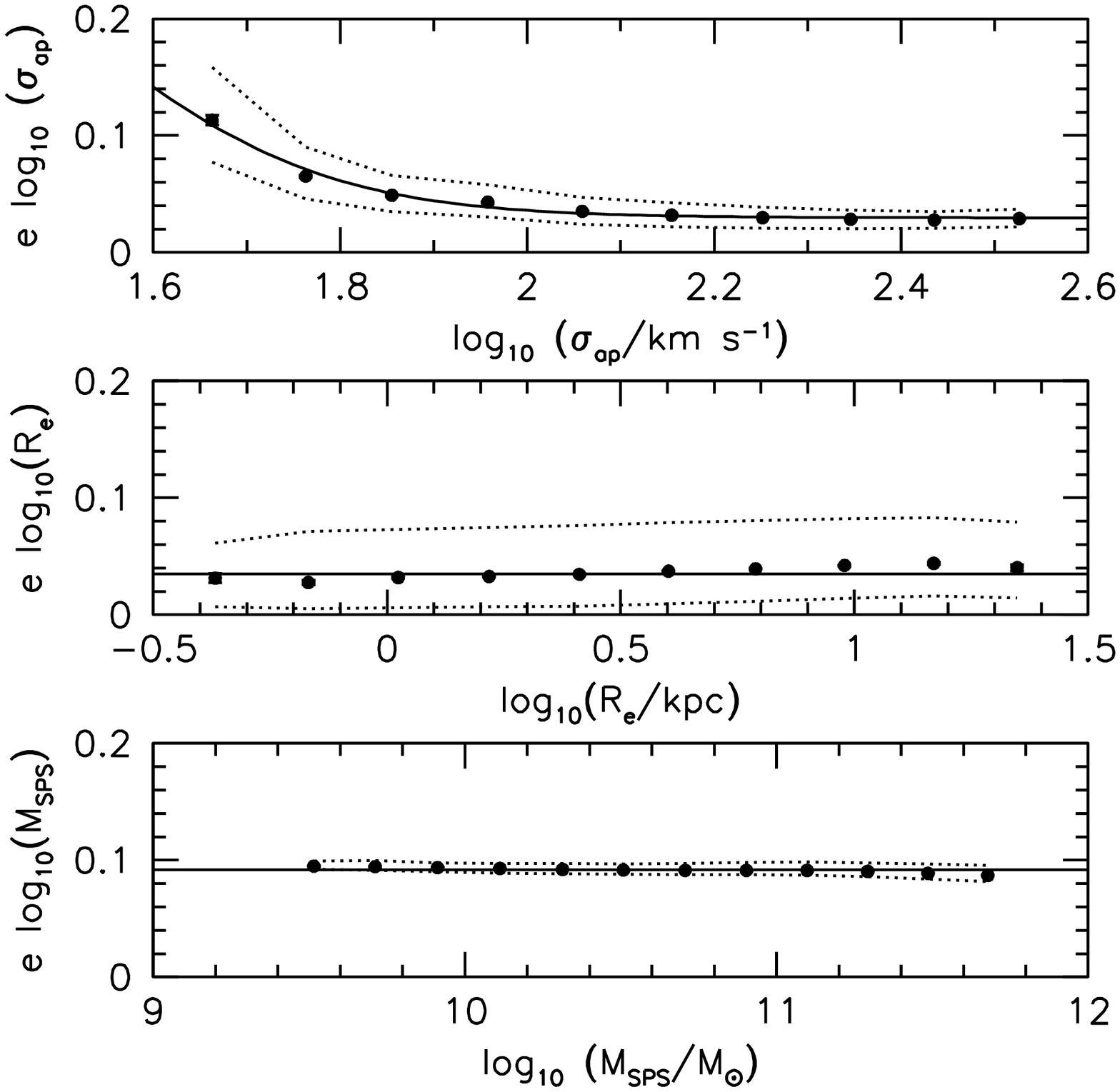,width=0.47\textwidth}
}
\caption{Measurement errors on velocity dispersion ($\sigma_{\rm
    ap}$), half-light size ($\Re$), and stellar mass
  ($\Msps$). Typical errors are 0.03 dex on velocity dispersion, 0.03
  dex on size, and 0.1 dex on stellar mass. The average errors are
  fitted with Eq.~\ref{eq:power2}, with parameters given in
  Table~\ref{tab:fits}.}
\label{fig:err}
\end{figure}

Additionally there is scatter in the aperture radius at fixed stellar
mass from the redshift sampling of galaxies, and is given in
Fig.~\ref{fig:rap}.  It turns out that this adds negligible scatter to
the VM relation, and so we do not include it in our fiducial models.

There are measurement errors in stellar masses, sizes and velocity
dispersions. These measurement errors are roughly constant
(Fig.~\ref{fig:err}) with $\sigma_{\log \Msps}\simeq 0.1$,
$\sigma_{\log R_{\rm e}}\simeq 0.035$ and $\sigma_{\log \sigma_{\rm
    ap}}\simeq 0.04$.  For our purposes the measurement errors in
stellar masses and sizes are more important than the measurement
errors in velocity dispersions. This is because we bin galaxies in
both stellar mass and size, but not velocity dispersion. Errors in
stellar masses can change the slope of the correlation between VM and
RM residuals, and thus are especially important to account for.

\label{lastpage}

\end{document}